\RequirePackage{fix-cm}
\documentclass[runningheads,a4paper]{llncs} 
\usepackage{graphicx}
\usepackage{natbib}
\usepackage{breakcites}
\usepackage{amsmath}
\usepackage{url}
\usepackage{placeins}
\usepackage[caption=false]{subfig}
\usepackage[a4paper, margin=1.5in]{geometry}
\usepackage[hidelinks]{hyperref}
\begin{document}
	
	\title{Brain Drain and Brain Gain in Russia: Analyzing International Migration of Researchers by Discipline using Scopus Bibliometric Data 1996-2020}
	
	
	\titlerunning{Brain Drain and Brain Gain in Russia}  
	
	\author{Alexander Subbotin \and Samin Aref}
	
	\authorrunning{Subbotin and Aref} 
	
	\institute{Lomonosov Moscow State University, ul. Leninskiye Gory, 1\\
		119991 Moscow, Russia\\
		\email{subbotin@demogr.mpg.de} 
		\and
		Laboratory of Digital and Computational Demography,\\
		Max Planck Institute for Demographic Research,\\ Konrad-Zuse-Str. 1, Rostock 18057, Germany\\
		\email{aref@demogr.mpg.de}\\}
	
	\date{}

	\maketitle
	
	\begin{abstract}
		We study international mobility in academia, with a focus on the migration of published researchers to and from Russia. Using an exhaustive set of over $2.4$ million Scopus publications, we analyze all researchers who have published with a Russian affiliation address in Scopus-indexed sources in 1996-2020. The migration of researchers is observed through the changes in their affiliation addresses, which altered their mode countries of affiliation across different years. While only $5.2\%$ of these researchers were internationally mobile, they accounted for a substantial proportion of citations. Our estimates of net migration rates indicate that while Russia was a donor country in the late 1990s and early 2000s, it has experienced a relatively balanced circulation of researchers in more recent years. These findings suggest that the current trends in scholarly migration in Russia could be better framed as brain circulation, rather than as brain drain. Overall, researchers emigrating from Russia outnumbered and outperformed researchers immigrating to Russia. Our analysis on the subject categories of publication venues shows that in the past 25 years, Russia has, overall, suffered a net loss in most disciplines, and most notably in the five disciplines of neuroscience, decision sciences, mathematics, biochemistry, and pharmacology. We demonstrate the robustness of our main findings under random exclusion of data and changes in numeric parameters. Our substantive results shed light on new aspects of international mobility in academia, and on the impact of this mobility on a national science system, which have direct implications for policy development. Methodologically, our novel approach to handling big data can be adopted as a framework of analysis for studying scholarly migration in other countries.
		\keywords{High-skilled migration \and Scholarly migration \and Brain circulation \and Digital demography \and Science of science \and Scientometrics
			\\
			\\
			Peer-reviewed and accepted author-copy. Publisher’s verified version:\\ Subbotin, A. \& Aref, S. (2021) Brain Drain and Brain Gain in Russia: Analyzing International Migration of Researchers by Discipline using Scopus Bibliometric Data 1996-2020. Forthcoming in \textit{Scientometrics}.
		}
	\end{abstract}
	
	\clearpage
	
	\section{Introduction}
	\label{s:intro}
	In an interconnected world, national science systems cannot be studied in a vacuum, while disregarding the impact of these systems on human mobility and migration. Today’s modern societies are knowledge societies \citep{lane1966decline,stehr2018modern}. Highly skilled specialists, including researchers, contribute to the consolidation of existing information, and to the dissemination of knowledge in various fields. As states compete for talented people, international migration can both strengthen and weaken individual countries in terms of their total human capital, which can, in turn, affect the socioeconomic and innovative development of these countries. The large increase in high-skilled migration between countries in recent years poses new challenges for both researchers and policy-makers.
	
	In this paper, we focus on Russia as a player in the global international migration system. Large shares of the Russian population are on the move for various reasons \citep{iontsev_emigration_2016}. Russia is also an attractive destination for some international migrants, especially migrants from the former Soviet republics \citep{bedrina_migration_2018,rasuly-paleczek_migration_2017}. Moreover, some migrants may use Russia as a transit stop for further migration to other countries \citep{rybakovsky_international_2005}.
	Previous studies have suggested that Russia is both a donor country and a recipient country \citep{di_bartolomeo_regional_2014,podolskaya_migration_2020} for migration among the general population. However, scholars who have taken the characteristics of migrants into account have argued that Russia is more of a donor country \citep{ushkalov__2001,zubova_human_2012,kolesnikova_current_2014}; i.e., that it is a country on the losing side of the international exchange of highly skilled individuals. 
	
	The numbers of published researchers in Russia and their outputs are perhaps not as well-known as those of other developed countries. According to SciVal 2010-2019 data\footnote{SciVal is a research profiling system and a web-based analytics solution provided by Elsevier. \url{www.scival.com} (accessed on 31/07/2020).}, Russia has over $440,000$ published researchers (comparable to Australia and Italy) who have produced nearly $700,000$ pieces of scholarly publications (comparable to South Korea). Despite its research contributions, Russia has remained a relatively under-studied case in the science of science and the high-skilled migration literature. Most studies on these topics have been limited to providing qualitative explanations for the emigration of specialists, which often do not go beyond suggesting the necessity of facilitating circular migration in Russia \citep{yurevich_global_2019,iontsev_problems_2017,molodikova_migration_2016,kolesnikova_current_2014,ryazantsev_russia_2013,volz_utechka_2002,ushkalov__2001,naumova_russias_1998,taylor_transformation_1996}. Therefore, a deeper analysis is needed that quantitatively examines the international movements of researchers in Russia, and their implications for different fields of science. According to previous studies, large numbers of scientists in mathematics \citep{volz_utechka_2002,ryazantsev_russia_2013}, physics \citep{ball_russian_2005,ryazantsev_russia_2013}, and computer science \citep{ryazantsev_russia_2013,antoshchuk_female_2018} leave Russia. The major destination countries for scholars from Russia appear to be the United States (US), Germany, France, the United Kingdom (UK), and Japan \citep{korobkov_russian_2012}. The movers tend to be from major scientific centers in Moscow, St. Petersburg, Novosibirsk, and Yekaterinburg. In addition, most of the movers come from younger age groups \citep{michael_beenstock_professor_raul_r_role_2015,iontsev_demographic_2015}, and thus had the potential to contribute to the Russian science system for long periods of time if they had stayed. Further research is needed to accurately quantify this phenomenon with respect to the similarities and the differences between researchers based on the types of mobility they are engaged in (non-movers, emigrants, immigrants, etc.); the origin and destination countries of migrant researchers; the interplay of researchers’ mobility patterns, levels of experience, and research performance; and the impact of migration on the Russian science system.
	
	Quantitative studies on the international migration of researchers are complicated by a lack of reliable, relevant, and comparable statistics. Recent studies on this topic have used bibliometric data to detect migrant populations among researchers, and to identify migration trajectories and flows for further analysis \citep{moed_bibliometric_2014,aref_demography_2019,kosyakov_impact_2019,gonzalez2020scholarly,zhao2021international}. This method involves tracking the international movements of researchers through the changes in their affiliation addresses. The feasibility of this approach has been tested in previous studies that estimated the migration flows of scholars between \citep{moed_bibliometric_2014,aref_demography_2019,kosyakov_impact_2019,zhao2021international}, or within countries \citep{gonzalez2020scholarly}. 
	
	There are also some studies specifically on the migration of Russian scholars that used bibliometric data. However, these studies were largely limited to studying the international migration of scientists affiliated with a specific institution \citep{sudakova_migration_2020,koksharov_international_2018}, or researchers from a particular scientific field \citep{malakhov_russian_2020,yurevich_brain_2018}.Recent analyses on Scopus bibliometric data have suggested that the largest outflow of researchers from one of the largest universities in Russia, the Ural Federal University (UrFU), occurred in the 1990s and early 2000s \citep{sudakova_migration_2020}; and that the main destinations of researchers moving from Russia have been the United States and Western European countries \citep{koksharov_international_2018}. It has also been shown that the main areas of scientific interest of scientists emigrating from the UrFU are the natural and the technical sciences \citep{koksharov_international_2018}. In another study that focused exclusively on migration in the field of mathematics, the authors analyzed data from the Web of Science for 2008-2018, and concluded that during this period, the international movements of most researchers were temporary \citep{malakhov_russian_2020}. Thus, there is support for the argument that these international movements of mathematicians should be referred to as ``brain circulation'', rather than as ``brain drain'' \citep{malakhov_russian_2020}. However, another study that focused on migration in the computer and information sciences provided support for the opposite argument by demonstrating that the large majority of researchers in these fields who emigrated have not returned to Russia \citep{yurevich_brain_2018}.
	
	Thus, the existing studies on the migration of Russian scholars have largely been limited to specific cases, and their findings are somewhat contradictory. In the present study, we adopt a comprehensive perspective by including all researchers who have published in Scopus-indexed sources with a Russian affiliation address at some point over the 1996-2020 period. We track the international movements of all such researchers in order to systematically analyze the impact of migration on the Russian science system overall, and in different fields of scholarship.
	
	\section{Materials and Methods}
	\label{s:data}
	\subsection{Scopus publications of all authors with ties to Russia}
	The availability of millions of publications in the Scopus database\footnote{Scopus is a database of peer-reviewed scientific literature \citep{falagas2008comparison,mongeon_journal_2016} covering $77$ million citable documents, $22\%$ of which are
		published in languages other than English \citep{scopus_coverage}.} allows us to study scholarly migration in Russia by aggregating the movements of each researcher who had affiliation ties to Russia at some point over the 1996-2020 period (up to the end of April 2020). The unit of data we use is an \textit{authorship record,} which is the linkage between an author affiliation and a publication. The data linked to an authorship record provide proxies not only for the geographic locations of researchers, but also for their research areas. Scopus annotates subject codes to more than $25,000$ indexed publication venues based on the topics they cover. This allows us to analyze the disciplines of internationally mobile researchers based on the subjects of their publications. There are $2,484,602$ publications in Scopus associated with over $659,000$ author profiles of researchers who have published with a Russian address at some point over the 1996-2020 period. These publications constitute the main dataset that we analyze.
	The two largest categories of these researchers were (1) those with only one publication (single-paper authors); and (2) those with multiple publications, but with no evidence of international mobility (non-movers). By contrast, just $5.2\%$ of these researchers were internationally mobile. However, the researchers in this small category were associated with $699,730$ of the publications in the dataset ($28\%$).
	
	\subsection{Data pre-processing}
	Examining Scopus author IDs \citep{kawashima_accuracy_2015,aman_does_2018} is the first step in identifying the authorship records of individual scholars, and, in turn, detecting mobility events. However, before movements can be detected, there are data quality issues with Scopus author IDs and affiliations \citep{gonzalez2020scholarly} that require some attention. Partly due to copyright, the affiliations are not standardized, and they may have substantially different formats. In a large majority of cases, the affiliation address has a country. However, there are $9,701$ authors in our dataset who have records in which no country is explicitly specified. These records come from $7,279$ distinct publications. Inspired by \cite{gonzalez2020scholarly}, we have chosen to use a neural network to predict the missing country information. The neural network takes the affiliation address of an authorship record and predicts the country associated with it. We used one million records for which the countries are specified as training data ($80\%$) and as test data ($20\%$). For $98.4\%$ of the test data, the neural network predicted the expected country. After ensuring that this method is highly accurate, we then used the trained neural network for predicting missing countries.
	
	There is evidence that the Scopus author identification system is reliable for analyzing the migration of researchers \citep{aman_does_2018}, as almost all author IDs correctly identify one researcher \citep{kawashima_accuracy_2015}. A \textit{precise} author ID is one that is only associated with publications of the same person. An evaluation of the accuracy of the system conducted in August 2020 showed that the precision of the Scopus author profiles was $98.3\%$\footnote{The recall of Scopus author profiles is also measured to be as high as $90.6\%$ \citep{loktev2020}.}\citep{loktev2020}. However, while the Scopus system is a gold-standard benchmark in the context of individual-level bibliometric data \citep{kawashima_accuracy_2015,aman_does_2018}, it remains an imperfect source of big data for analysis at the level of authors. Therefore, further pre-processing of the Scopus data is needed to accurately identify individual authors.
	
	Due to imperfections in the Scopus author identification system, $1.7\%$ of the author IDs could be associated with publications of more than one individual (possibly multiple individuals with the same name). We approach this problem by treating suspicious outliers in order to improve the reliability of author IDs. We apply an author disambiguation process \citep{gonzalez2020scholarly,dangelo_collecting_2020} to the authorship records that are more likely to be affected by the precision flaws in the Scopus author identification system. Authorship IDs that exceed either of the following two thresholds are deemed suspicious, and will thus be treated by our author disambiguation method. This method is conservative by design, as it assumes that there are distinctions, and searches for evidence of similarities. Author IDs are deemed suspicious if the values for the number of countries and the number of publications are extreme. The first threshold is being associated with more than six countries of affiliation. The second threshold is being associated with more than 292 publications (an average of more than one publication per month over a period of 24 years and 4 months).
	
	Among more than $659,000$ distinct author IDs in our data, $3,563$ author IDs (less than $0.5\%$) are deemed suspicious. These author IDs are associated with $334,484$ distinct publications. We disambiguate these records using an unsupervised machine learning algorithm \citep{gonzalez2020scholarly} inspired by the state-of-the-art methods proposed in the literature \citep{dangelo_collecting_2020}, and assign revised author IDs using the following method. Our author disambiguation algorithm makes pairwise comparisons between every two records with the same author ID, and allocates scores that are higher if the two authorship records have similar traits, and that are lower if the records are dissimilar. Then, the scores are summed up and a distance matrix is produced for all pairs of authorship records with the same author ID. Using agglomerative clustering from the scikit-learn package in Python \citep{pedregosa_scikit-learn_2011}, we obtain clusters of highly similar authorship records. Finally, a revised author ID is issued to each cluster \citep{gonzalez2020scholarly}. After this author disambiguation method is implemented, revised author IDs can be issued to the subset of suspicious authorship records. Note that this process is not meant to increase the precision level of Scopus author IDs to $100\%$. Instead, it is designed to reduce the possible impact of outliers (which might have resulted from the precision flaws of the Scopus author IDs) on migration estimates.
	
	\subsection{Fields and subfields of scholarship}
	According to the All Science Journal Classification (ASJC), there are four major fields of science: \textit{life sciences} (which includes five subfields\footnote{The life sciences include
		(1) the agricultural and biological sciences;
		(2) biochemistry, genetics and molecular biology;
		(3) immunology and microbiology;
		(4) neuroscience; and
		(5) pharmacology, toxicology and pharmaceutics.}), \textit{social sciences} (which includes six subfields\footnote{The social sciences include
		(1) the arts and humanities;
		(2) business, management and accounting;
		(3) decision sciences;
		(4) economics, econometrics and finance;
		(5) psychology; and
		(6) other social sciences.}), \textit{physical sciences} (which includes ten subfields\footnote{The physical sciences include
		(1) chemical engineering;
		(2) chemistry;
		(3) computer science;
		(4) earth and planetary sciences;
		(5) energy;
		(6) engineering;
		(7) environmental science;
		(8) materials science;
		(9) mathematics; and
		(10) physics and astronomy.}), and \textit{health sciences} (which includes five subfields\footnote{The health sciences include
		(1) medicine;
		(2) nursing;
		(3) veterinary;
		(4) dentistry; and
		(5) health professions.}).
	Each publication venue in Scopus may be classified by multiple ASJC codes, which determine the fields and subfields of the topics they cover. At the level of the four major ASJC fields of science, we consider that a given researcher belongs to either one of the four fields, or to a fifth \textit{multidisciplinary} group.
	
	We initially compute the frequency $f_m$ of each of the four major fields $m\in\{\text{physical,}$ $\text{health, social, life}\}$ for the authorship records of each researcher. For each researcher, we compare $f_m$ with the mean, $\mu_m$, and standard deviation, $\sigma_m$, of frequencies for field $m$ among all researchers in our dataset. This comparison involves calculating four Z-scores for each researcher using $Z_m=(f_m-\mu_m)/\sigma_m$. Based on the largest Z-score that exceeds the threshold of $\alpha=1$, we group the researchers into one of the four groups of \textit{health, life, physical,} or \textit{social} sciences\footnote{The value of $\alpha$ is selected such that only $10\%$ of researchers become multidisciplinary. Stricter limits based on a larger $\alpha$ lead to clearer boundaries between the four main fields, and to more individuals belonging to the multidisciplinary group.}. For $10\%$ of researchers, neither of the Z-scores exceeded the threshold of $\alpha=1$, and we group them as \textit{multidisciplinary}.
	
	\subsection{Detecting moves and mobility types of researchers}
	For analyzing scholarly migration, we borrow well-known and fundamental concepts such as origin, destination, and migrant from migration studies, and repurpose them for usage in the context of academic migration. Accordingly, a country of \textit{academic origin}\footnote{Note that the country of academic origin is not meant to be a proxy for nationality or place of birth, but rather refers to the country that is likely to have invested in the professional development of the person who becomes a published researcher in that country.} is the mode country of affiliation for the publications during the first year of publishing. The \textit{academic destination} country is determined by the mode country of affiliation in the latest year of publishing. To refer to a researcher who has an international mobility event, we use the term academic migrant (or migrant, for brevity). We consider an international mobility event if the changes in affiliations across two different years are such that the mode of country of affiliation changes for a researcher\footnote{For a migration event to be considered when there are multiple modes in a given year, the change should be such that the previous mode country disappears from the list of new mode countries.}. 
	
	We define four categories for academic migrants based on their countries of academic origin and destination. In our analysis based on the whole 1996-2020 time period, each published researcher belongs to one of six of the following categories (migrants belong to one of the last four categories):\\
	(1) \textit{Single-paper author} (a researcher with a single publication),\\
	(2) \textit{Non-mover} (a researcher with multiple publications, but with no evidence of international mobility),\\
	(3) \textit{Immigrant} (origin: not Russia, destination: Russia),\\
	(4) \textit{Emigrant} (origin: Russia, destination: not Russia),\\
	(5) \textit{Return migrant} (origin: Russia, destination: Russia, with international migration),\\
	(6) \textit{Transient} (having Russia as a mode country, but not as an origin or as a destination).
	
	\subsection{Quantifying contributions of researchers by subfield}
	\label{ss:quantify}
	At the level of the 26 subfields (disciplines) of the ASJC, we consider that researchers may have been active in and contributing to several of them. Therefore, we define and use the concept of \textit{normalized contribution} to quantify the contribution of a given researcher to different fields in a normalized way. The normalized contribution $NC^j_{(d)}$ of researcher $j$ (among a total of $k$ researchers) in discipline $d$ (among a total of $n$ disciplines) is defined and formulated in Eq.\ \eqref{eq:NC} based on the relative frequency of discipline $d$ in his/her authorship records. $s^j_d$ is the frequency of discipline $d$ in the authorship records of individual $j$. The denominator in Eq.\ \eqref{eq:NC} is the sum of frequencies of $n$ disciplines in the authorship records of individual $j$.
	\begin{equation}
		\label{eq:NC}
		NC^j_{(d)}=\frac{s^j_d}{\sum_{i=1}^{n} s^j_i} \quad
		i = 1, \dots, n \quad \forall j \in \{1,\dots, k\}
	\end{equation}
	
	Table \ref{tab2} provides an illustrative example of the authorship records of a researcher (with author ID $x$) who is associated with a total of three distinct publications (distinguished by Digital Object Identifiers -- DOIs) in multiple disciplines. This researcher is classified as an emigrant, since her mode country of affiliation in 2012 was Russia, while her mode country in 2015 was the US. According to Eq.\ \eqref{eq:NC}, the normalized contributions of the researcher are $$NC^x_{(\text{chem.})}={2}/{5}, NC^x_{(\text{energy})}={1}/{5}, NC^x_{(\text{maths})}={2}/{5}.$$
	
	\begin{table}[hb!]
		\caption{Example of fictitious authorship records with multiple subjects and countries}
		\label{tab2}
		\centering 
		\begin{tabular}{lllll}
			\hline 
			ID & DOI & ASJC Subject   & Country & Year \\ \hline
			x  & 111 & Maths   & Russia & 2012 \\
			x  & 222 & Chemistry, Energy  & Russia, US & 2013 \\
			x  & 333 & Maths, Chemistry & US  & 2015 \\ \hline
		\end{tabular}\\[9pt]
	\end{table}
	
	To aggregate the normalized contributions of migrants in a given discipline, the \textit{normalized count} $P_d$ of migrants in discipline $d$ is defined, which is calculated by adding up all the normalized contributions of migrant researchers, as formulated in Eq.\ \eqref{eq:normalizedcount}. $P_d$ can be thought of as a weighted count of internationally mobile researchers in discipline $d$. This value is normalized by giving fractional weights to individuals based on how active they are in discipline $d$ compared to in their other disciplines.
	
	\begin{equation}
		\label{eq:normalizedcount}
		P_d = \sum_{j=1}^{k} NC^j_{(d)}
	\end{equation}
	Given that each mobile researcher belongs to one of the four categories of migrants, the normalized count can be similarly computed based on the normalized contributions of each type of migrant. Accordingly, we obtain $P^{\text{imm}}_d$,$P^{\text{emi}}_d$,$P^{\text{ret}}_d$, and $P^{\text{tra}}_d$ respectively (with specific $k$ values for each category of migrants) as normalized populations of immigrants, emigrants, return migrants, and transients in discipline $d$. In the next section, we demonstrate how these new quantities can be used for analyzing scholarly migration.
	
	\section{Results}
	\label{s:results}
	
	We present the main results of our analysis in this section. First, we explore the citation-based performance of researchers by their mobility type. Next, we discuss the geography of mobile researchers, and analyze their migration flows. We then compare the performance of immigrants and emigrants based on the citations they received (normalized by years of experience and fields, and disaggregated by countries). Next, we present our estimates of migration rates in order to evaluate brain circulation in the aggregate, and by major fields. In the last part of the results section, we evaluate the overall impact of migration on each scientific discipline in Russia.
	
	\subsection{Citation-based performance of researchers by mobility type}
	
	For all researchers in our dataset, we link the authorship records with citation data from Scopus to obtain an individual-level measure of citation performance. To compare the citation-based performances of researchers with different levels of experience, we use \textit{academic age} \citep{aref_demography_2019}, which equals the number of years since the first publication. We calculate an \textit{annual citation rate} by dividing a researcher’s total number of citations (as of April 2020) by his/her academic age. The results are provided in Table \ref{tab:citations_imm_em}.
	
	\begin{table}[b]
		\caption{Average and standard deviation of annual citations by field and mobility type}
		\label{tab:citations_imm_em}
		\centering
		\resizebox{\textwidth}{!}{%
			\begin{tabular}{p{2.1cm}llllllr}
				\hline
				& Life sci.    & Social sci.    & Physical sci.    & Health sci.     & Multidisc. & $\% ^{\dagger}$ \\ \hline 
				Immigrant  & $15.6 \pm 44.5$  & $3.4 \pm 11.1$   & $29.3 \pm 210.3$    & $10.4 \pm 48.0$    & $18.7 \pm 55.5$ & $1.4 \%$\\
				Emigrant  & $37.9 \pm 98.6$  & $3.1 \pm 10.6$   & $52.8 \pm 263.3$    & $25.1 \pm 115.1$   & $42.2 \pm 153.1$ & $2.3 \%$\\
				Returner & $24.2 \pm 51.1$  & $3.6 \pm 6.6$    & $58.9 \pm 272.2$    & $24.1 \pm 122.6$   & $36.3 \pm 87.6$ & $0.9 \%$ \\
				Transient  & $50.0 \pm 130.3$  & $14.2 \pm 45.8$   & $56.0 \pm 249.8$    & $31.2 \pm 60.0$    & $40.8 \pm 92.8$ & $0.6 \%$ \\
				Non-mover  & $4.7 \pm 17.7$   & $1.3 \pm 5.1$    & $10.0 \pm 111.4$    & $3.0 \pm 25.2$    & $7.1 \pm 35.5$ & $41.0 \%$ \\ 
				1-paper author & $0.4 \pm 1.2$  & $0.1 \pm 0.5$    & $0.3 \pm 4.2$    & $0.3 \pm 3.0$    & $0.5 \pm 3.5$  & $49.2 \%$\\ 
				\hline  
			\end{tabular}
		}
		$\quad \dagger$ The right-most column shows the frequencies of researchers for each mobility type.
	\end{table}
	
	In Table \ref{tab:citations_imm_em}, the averages and standard deviations of annual citation rates are provided for different groups by fields. Single-paper authors are separated from non-movers to allow for better comparisons between researchers who were exposed to mobility. The results show that there were substantial disparities in citation performance between migrants and non-movers across major fields (despite separating single-paper authors from non-movers). The average citations of non-movers were substantially lower than the average citations of emigrants and citations of immigrants. As expected, the annual citations were lowest for the single-paper authors. The disparities in citation-based performance suggest that while internationally mobile researchers made up only a small fraction of all researchers in the sample ($5.2\%$), they were making essential contributions to the Russian science system, as indicated by the substantial numbers of citations they received when their fields and years of experience are taken into consideration. Note that summing up the percentages in the right-most column of Table \ref{tab:citations_imm_em} gives $95.4\%$. The remaining $4.6\%$ researchers we have excluded from the study on the basis of their mode countries, which suggests that they are not relevant to this study. These researchers did not have Russia as a mode country in any given year, despite having at least one Russian affiliation in their publications.
	
	The results displayed in Table \ref{tab:citations_imm_em} also reveal considerable differences in the numbers of citations received by immigrants and emigrants across different fields. We observe that the average citation rates were generally higher for the emigrants than for the immigrants, except in the social sciences, where the average citation rates of immigrants and emigrants were somewhat similar. These findings suggest that in most major fields, internationally mobile researchers who came to Russia performed worse than those who left Russia in terms of the citations they received, while controlling for the differences in years of academic experience and fields. As the standard deviations in Table \ref{tab:citations_imm_em} are shown to be large, we later analyze the differences in the citation-based performance levels of immigrants and emigrants using a different method in Figure \ref{fig:Emi_vs_Imm_cited}.
	
	
	Among the major fields of scholarship, there were substantial differences in the average rates of citations received by the researchers. This pattern was consistent with our general expectations \citep{Leydesdorff_citation, marmolejo2015mobility,bedenlier2017internationalization,horta2019mobility}. Multidisciplinary researchers had the highest average rate at $9.2$ citations per year, followed by physical science researchers at $6.5$ citations per year, life science researchers at $4.3$ citations per year, health science researchers at $1.9$ citations per year, and social science researchers at $0.6$ citations per year. We make use of these values to produce field-normalized citation rates later in this section. These differences between fields can also be seen by comparing the different columns of Table \ref{tab:citations_imm_em}.
	
	\subsection{Flows, origins, and destinations}
	\label{ss:origin}
	
\begin{figure*}
		\centering
		\includegraphics[angle=90,origin=c,width=0.74\textwidth]{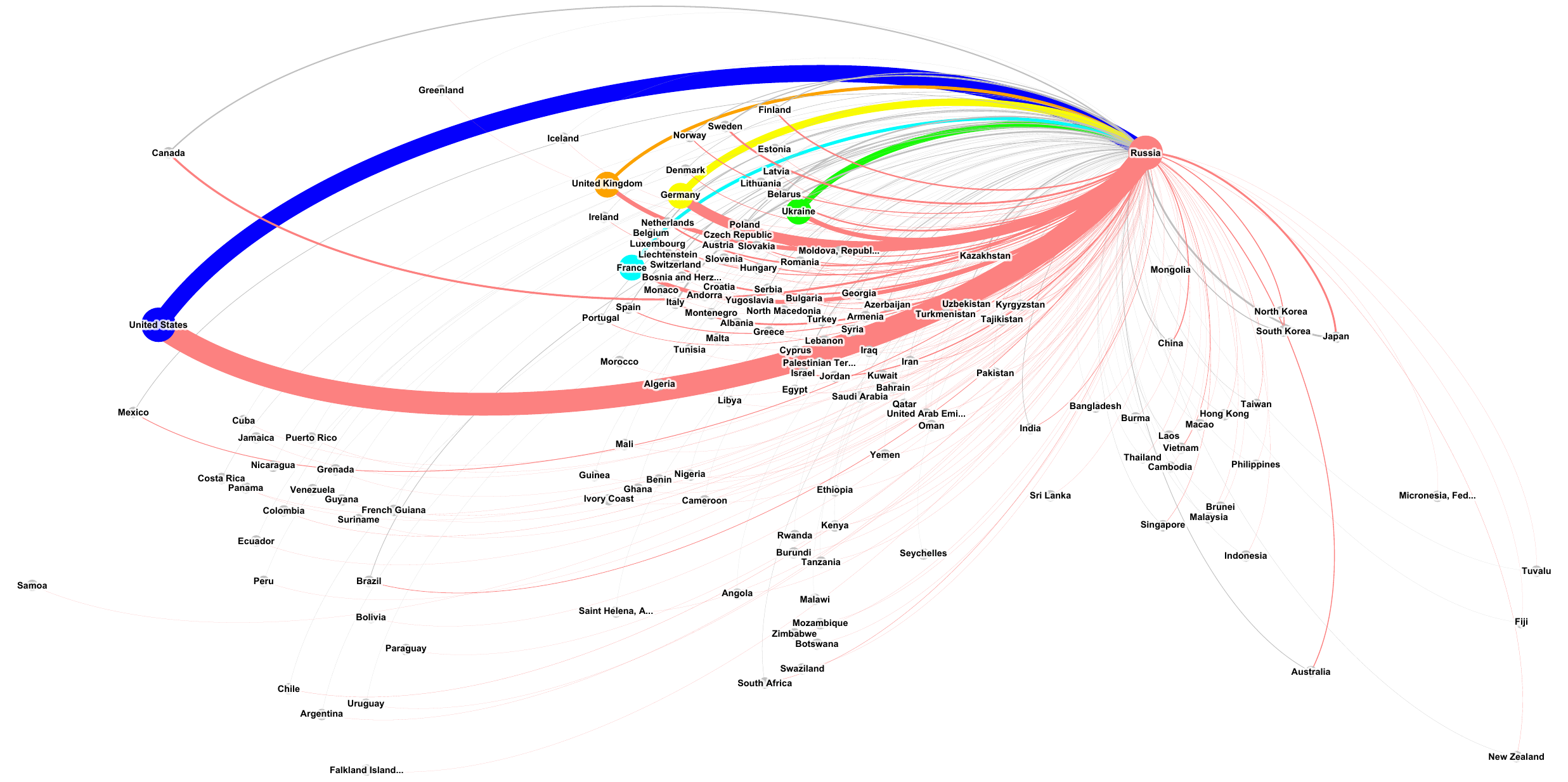}
		\caption{Network of movements to and from Russia among researchers over the 1996-2020 period. Directions of edges are clockwise. Common origins and destinations are shown with distinct colors. Colors of the flows are based on the origin country. Thickness of an edge is proportional to the flow it represents. See the figure on the screen for high resolution.} \label{fig:network}
	\end{figure*}

	Figure \ref{fig:network} illustrates the international paths for researchers who moved to or from Russia over the 1996-2020 period. It shows that the US and Russia are connected by two edges (whose directions are clockwise): blue (moves from the US to Russia) and pink (moves from Russia to the US). The pink edge is thicker than the blue one, which means that the moves from Russia to the US outnumbered the moves in the opposite direction. 
	
	From a Russian perspective, the US, Germany, the UK, and France were more likely to be destinations than to be origins, which suggests that the flows of immigrants and emigrants between Russia and these countries were imbalanced. The five most common countries of academic origin for immigrants were the US, Ukraine, Germany, France, and the UK. As destinations for emigrants, the US was again the most common country, followed by Germany, the UK, Ukraine, and France. Moreover, the total emigration flows to these five most frequent destinations was more than $50\%$ greater than that of the immigration flows from the five most frequent origins.
	
	An exception among the top countries of origin was Ukraine, which was sending twice as many immigrants to Russia as the number of emigrants it was receiving from Russia. In the general population, the migration relationship between Russia and Ukraine has been close, with Russia long being the most common destination for Ukrainian migrants \citep{cipko_contemporary_2006,mukomel_migration_2017}. In addition, from April 2014 to February 2016, more than one million people migrated from Ukraine to Russia following the 2014 revolution in Ukraine \citep{molodikova_migration_2016}.
	
	Over the 1996-2020 period, Canada, Finland, Sweden, the Netherlands, China, and Kazakhstan received larger flows of published researchers from Russia than they sent to Russia. A similar, though less distinct pattern can be observed for Spain, Italy, and South Korea. For Canada, Finland, and Sweden, the flows coming from Russia were around two and a half times larger than the flows in the opposite direction. Though it was not among the top 15 origin countries, Switzerland was a common academic destination country, receiving almost four times as many published researchers from Russia as it was sending to Russia. By contrast, Belarus, Uzbekistan, and Poland were more likely to be countries of origin rather than of destination. The migration patterns of academics from Belarus and Uzbekistan might be explained by the historical trend in patterns of general migration, whereby Russia has long been a primary destination for migrants from these countries \citep{titarenko_republic_2016,bedrina_migration_2018}.
	
	The rankings of the most common origins and destinations for transient scholars almost matched the rankings of the most common countries for immigrants and emigrants. The numbers of outgoing transients from Russia to China, Sweden, Poland, the US, Kazakhstan, Switzerland, and the UK were higher than the respective numbers of incoming transients from these countries to Russia.
	
	For return migrants, we look at the intermediate country(-ies) that a given return migrant was affiliated with while s/he was temporarily away from Russia\footnote{For a return migrant with multiple intermediate countries, we use equal weights adding up to one in measuring the frequency of intermediate countries.}. The five most common intermediate countries for return migrants were the US, Germany, France, the UK, and Ukraine.  
	
\begin{figure}
		\centering
		\subfloat[Health sciences]{
			\includegraphics[width=0.655\textwidth]{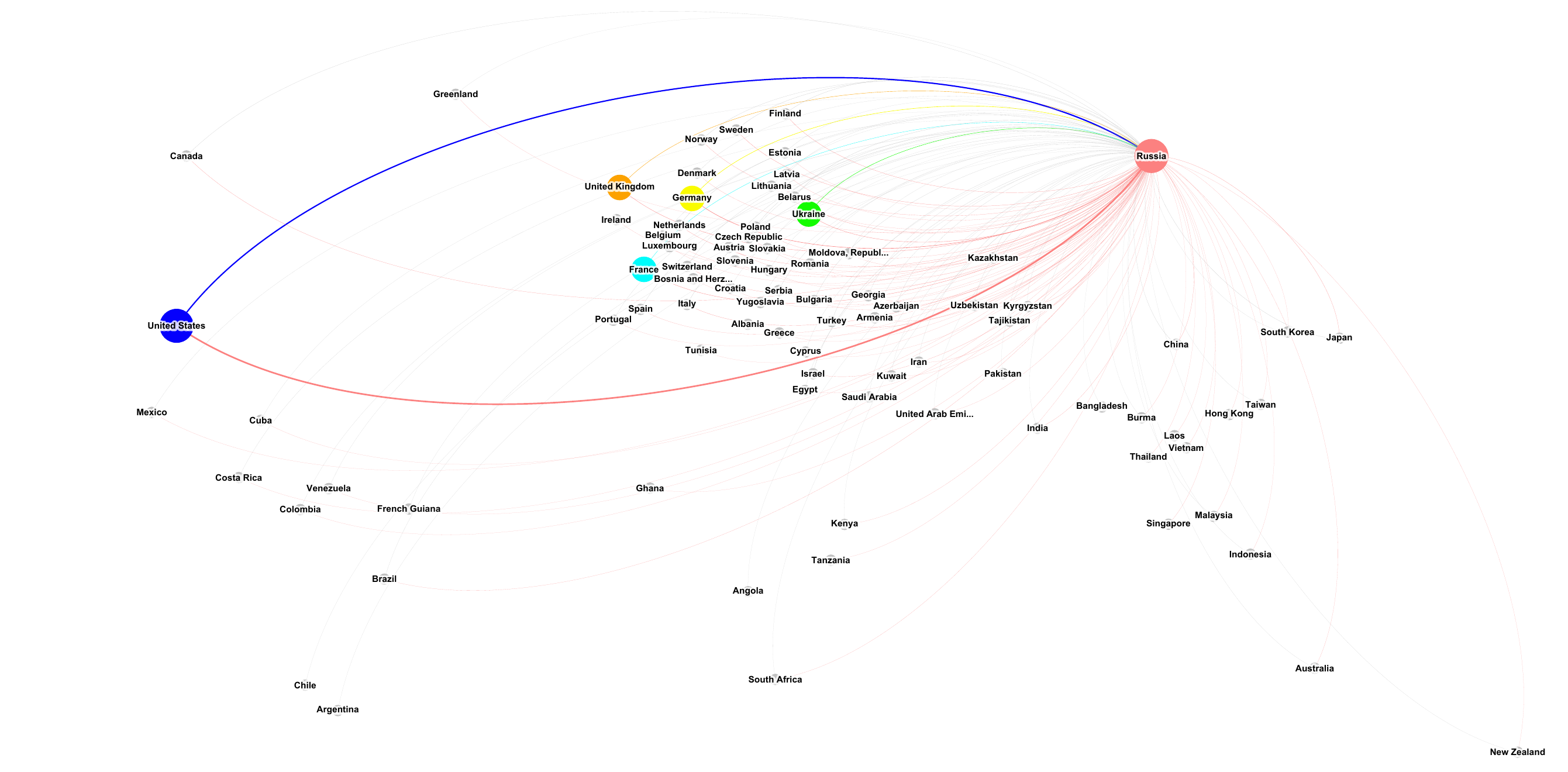}
			\label{subfig:health}
		} \hfill
		\subfloat[Life sciences]{
			\includegraphics[width=0.655\textwidth]{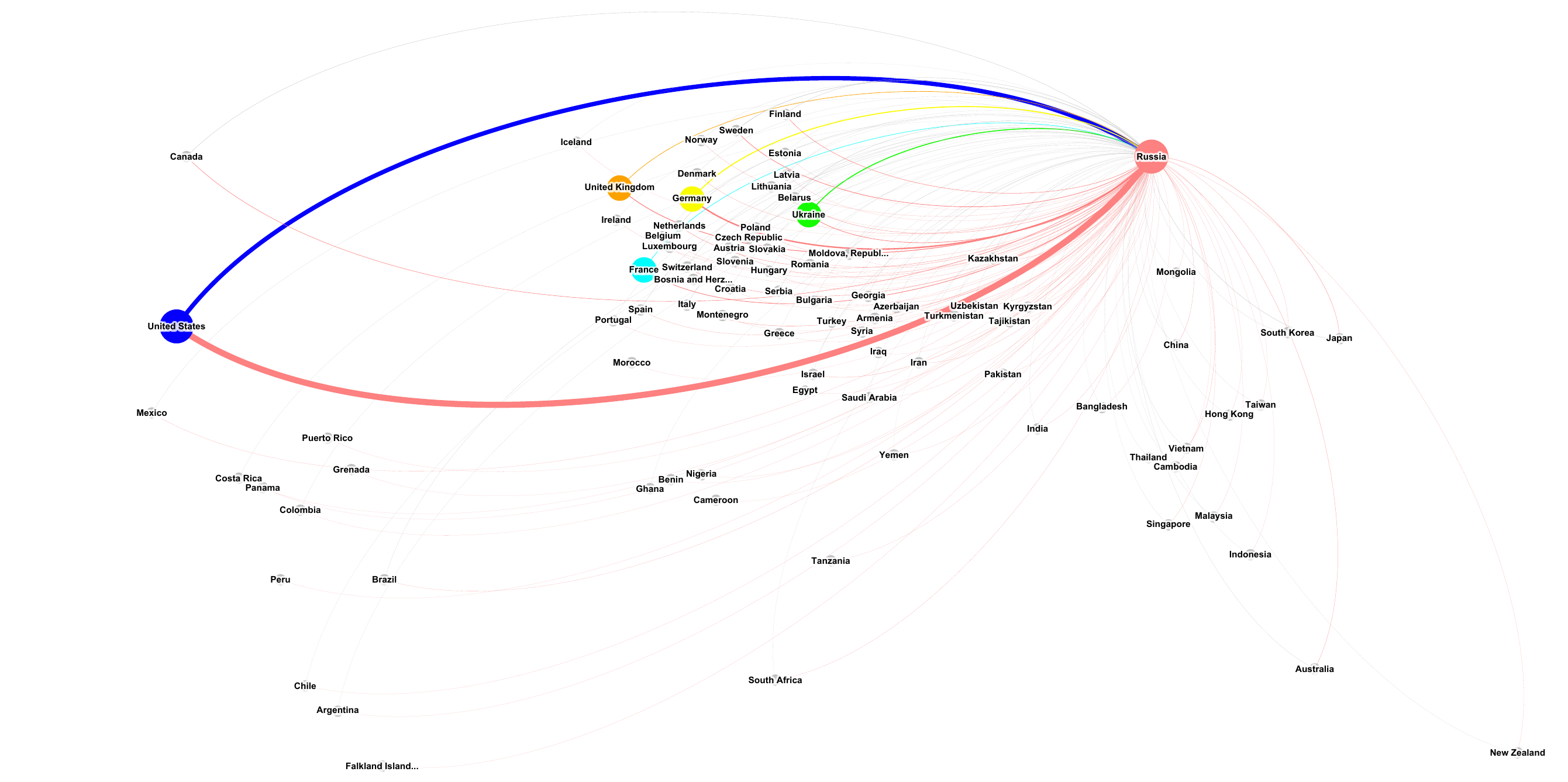}
			\label{subfig:life}
		} \hfill
		\subfloat[Physical sciences]{
			\includegraphics[width=0.655\textwidth]{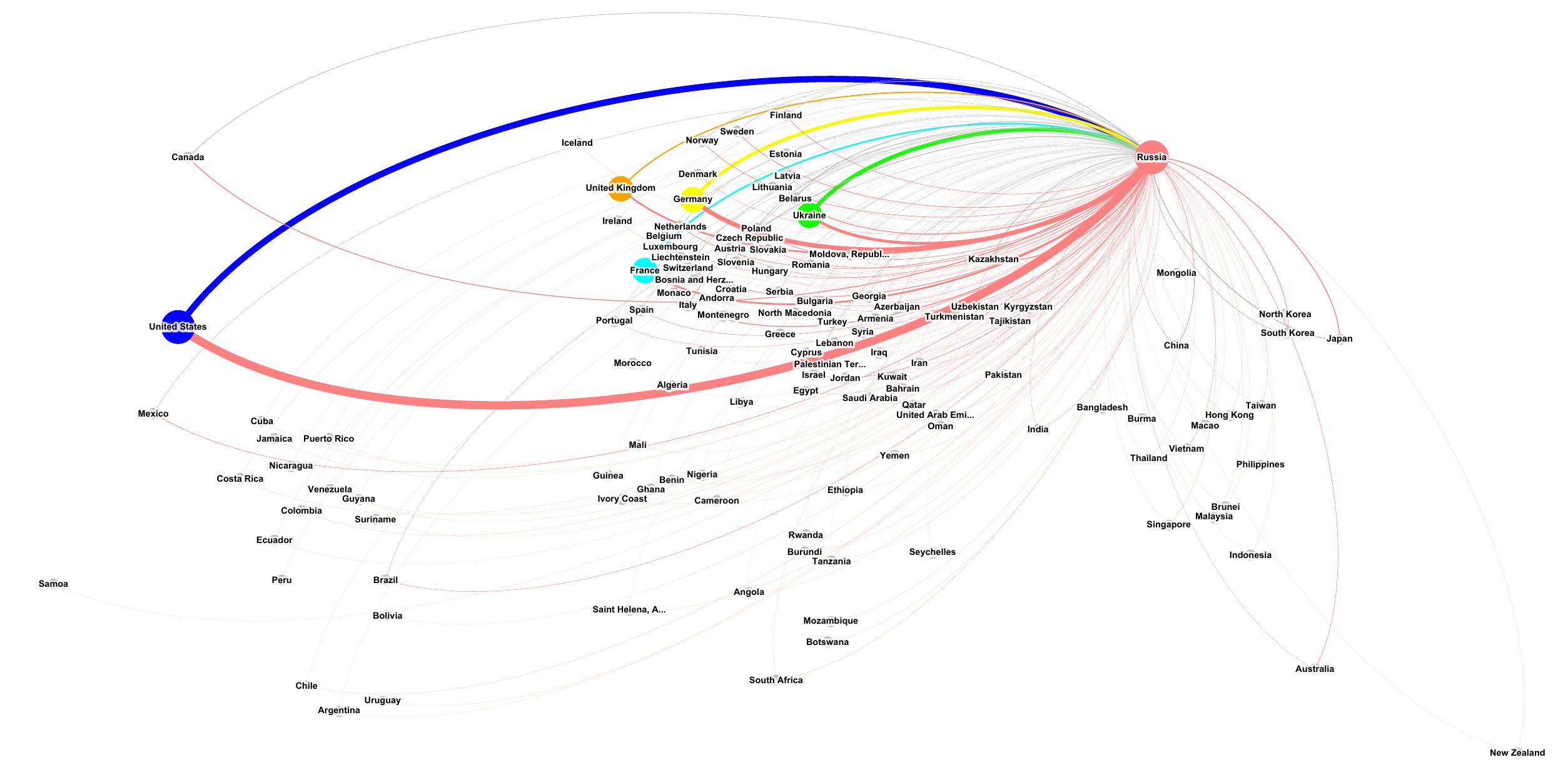}
			\label{subfig:physical}
		} \hfill
		\subfloat[Social sciences]{
			\includegraphics[width=0.655\textwidth]{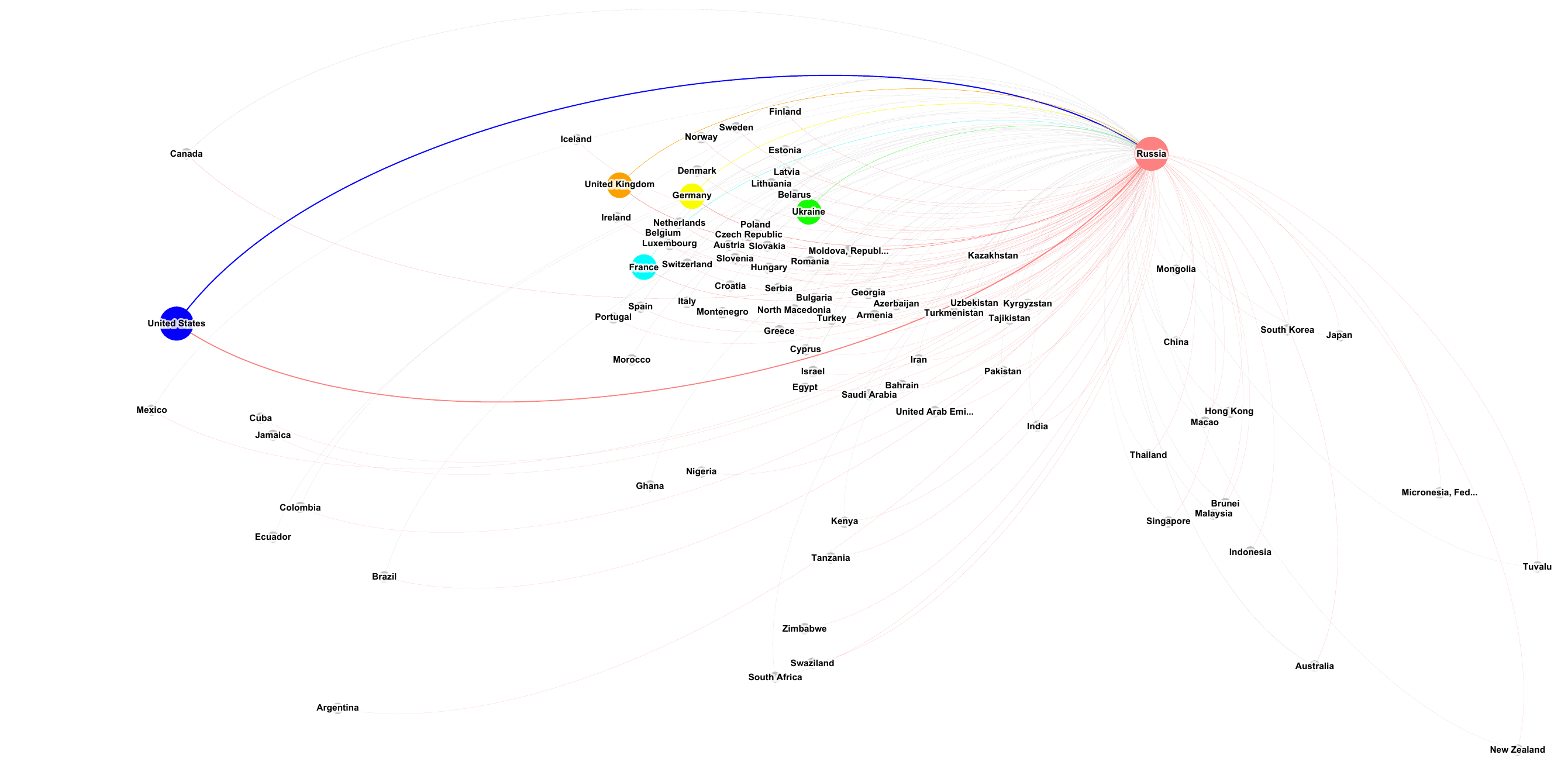}
			\label{subfig:social}
		}
		\caption{Migration flows among researchers in four major fields. Colors of the flows are based on the origin country. See the figure on the screen for high resolution.} \label{fig:networks_by_field}
	\end{figure}

	The international movements previously shown in Figure \ref{fig:network} are disaggregated based on the major field of the migrant researcher in Figure \ref{fig:networks_by_field}. The physical sciences had the largest total flow (22,376 moves by 15,308 researchers), followed by the multidisciplinary fields (13,670 moves by 9,013 researchers; which is not visualized), the life sciences (8,553 moves by 5,867 researchers), the health sciences (3,276 moves by 2,438 researchers), and the social sciences (2,379 moves by 1,824 researchers). Figure \ref{fig:networks_by_field} shows that the top five destination countries were the US, Germany, the UK, Ukraine, France, Kazakhstan, and Sweden; although their order varied depending on the major field of science. The US was consistently the most common destination. For the physical and the health sciences, the second- to the fifth-most common destinations were Germany, Ukraine, France, and the UK. For the social sciences, the second- to the fifth-most common destinations were the UK, Germany, Kazakhstan, and Ukraine.
	
	For the life sciences, the second- to the fifth-most common destinations were Germany, the UK, France, and Sweden. For Ukraine, the flows to Russia were larger than the flows from Russia in all four major fields. With the exception of Ukraine, the flows from Russia were larger than the flows in the opposite direction for most other countries and field combinations.
	
	\subsection{Citation-based performance of immigrants and emigrants by country}
	\label{ss:citation}
	
\begin{figure}
		\centering
		\subfloat[Composition of citation class by origin countries for immigrants]{
			\includegraphics[width=0.47\textwidth]{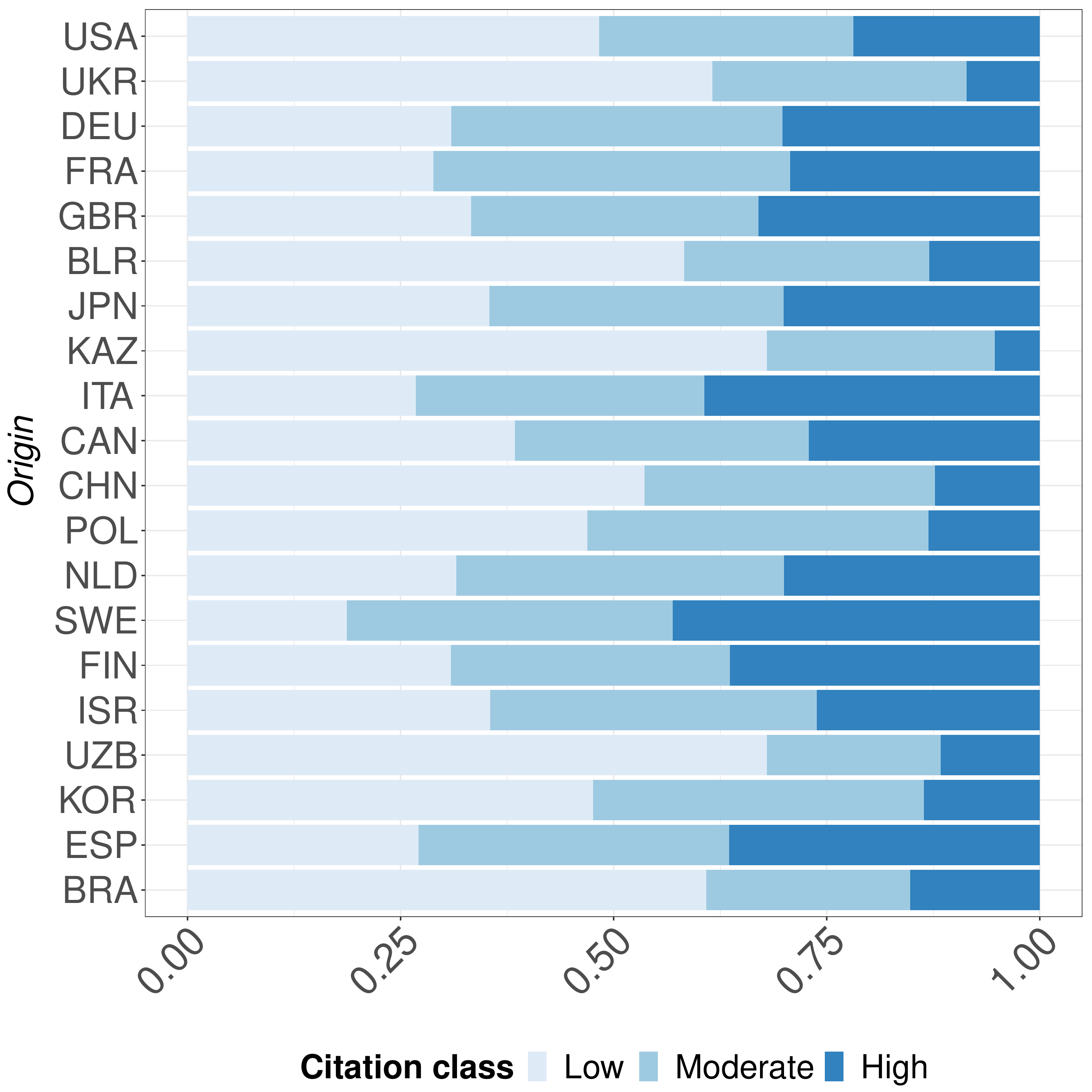}
			\label{subfig:immigrants_origin_LC}
		} 
		\hfill
		\subfloat[Composition of citation class by destination countries for emigrants]{
			\includegraphics[width=0.47\textwidth]{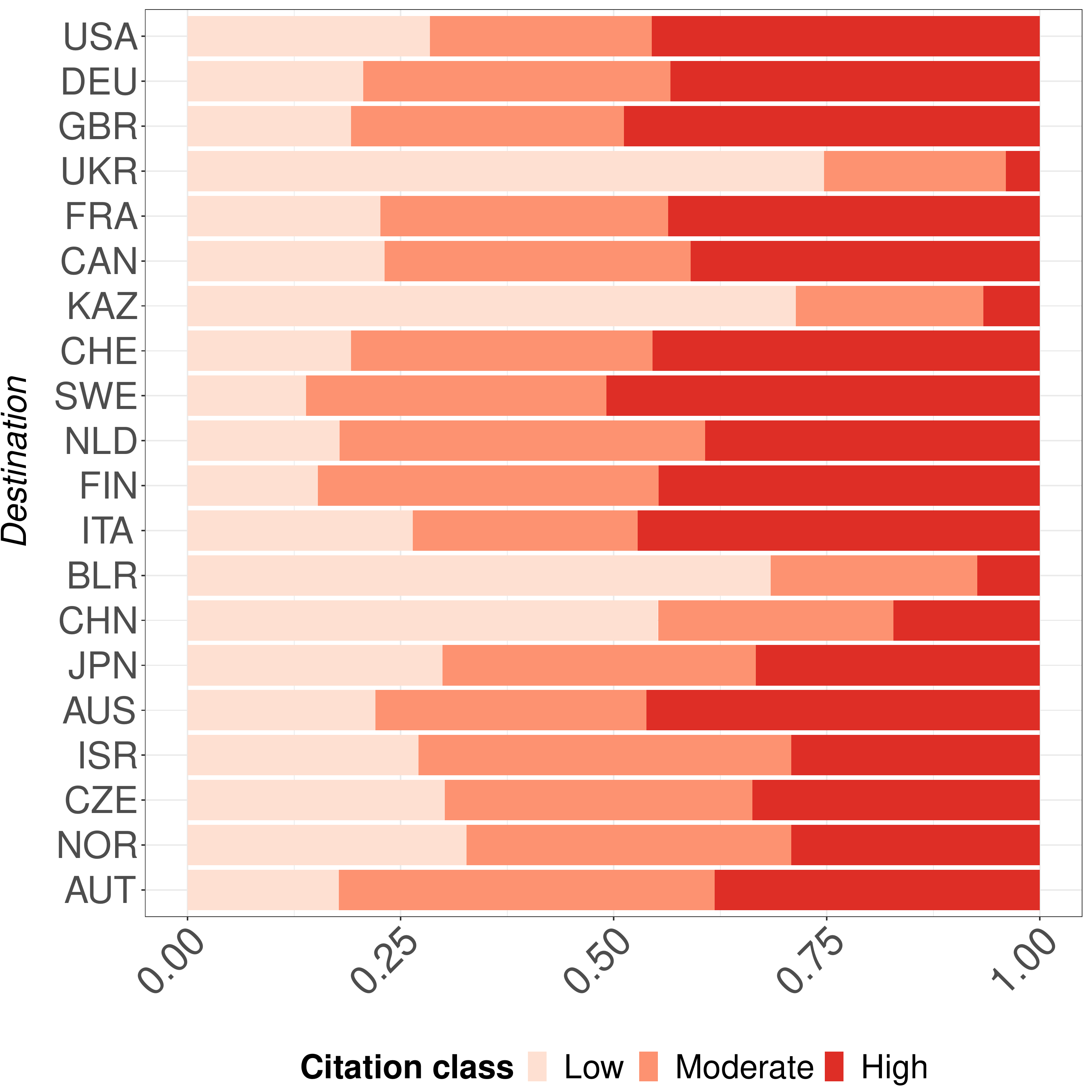}
			\label{subfig:emigrants_destination_LC}
		} 
		\caption{Composition of citation performance of top 20 origins for immigrants (a), and top 20 destinations for emigrants (b)} \label{ris:immigrants_emigrants_cited}
	\end{figure}

	We also look at the common origins and destinations of the researchers while taking into account their citations and academic age. To obtain a suitable citation-based measure of performance for migrants, we divide the annual citation rate of a researcher by the average rate in his/her field among migrants. After normalization, we define three groups of migrants based on the two 3-quantiles of the resulting distribution. \textit{Lowly cited migrants} are migrant researchers who had a field-normalized annual citation rate of less than 0.23 (the first tertile), while \textit{moderately cited migrants} are those who had a field-normalized annual citation rate of between $0.23$ and $2.10$ (i.e., between the first and second tertiles), and \textit{highly cited migrants} are those who had a field-normalized annual citation rate above $2.10$ (the second tertile). Figure \ref{ris:immigrants_emigrants_cited} shows the citation-based performance of migrant researchers from the most common origins for immigrants and the most common destinations for emigrants in descending order of the number of migrants.
	
	Figure \ref{ris:immigrants_emigrants_cited} shows that the US was both the most common origin of immigrants and the most common destination of emigrants. Focusing on the US as an example, we can see that $48.3\%$ of academic immigrants from the United States to Russia belonged to the lowly cited class, while $29.8\%$ were moderately cited, and $21.9\%$ were highly cited (see Subfigure \ref{subfig:immigrants_origin_LC}). These shares can be compared against the average migrant, who had an equal likelihood (of $33.3\%$) of belonging to any of the three classes. By contrast, for emigrants from Russia to the US, $45.5\%$ were in the highly cited category, $26.0\%$ were in the moderately cited category, and $28.4\%$ were in the lowly cited category (see Subfigure \ref{subfig:emigrants_destination_LC}). Taken together, these patterns show that there were substantial differences in the citation performance of migrants depending on the direction of the migration flow.
	
	Large proportions of scholars associated with Ukraine, Belarus, and Kazakhstan (as either origins or destinations) belonged to the low citation group. For example, $68.0\%$ of immigrants from Kazakhstan and $71.4\%$ of emigrants to Kazakhstan belonged to the low citation class. Ukraine received the highest proportion of low citation emigrants ($74.7\%$), and $61.6\%$ of the immigrants it sent to Russia were in the low citation group. Turning to the UK, we see that among immigrants, the three citation groups were almost equal in size; while among emigrants, larger shares belonged to the higher than to the lower citation groups. Comparing Subfigure \ref{subfig:immigrants_origin_LC} and Subfigure \ref{subfig:emigrants_destination_LC}, we can see that immigrants were more likely to be lowly cited, while emigrants were more likely to be highly cited; confirming the general pattern from Table \ref{tab:citations_imm_em}, but at a more fine-grained scale.
	
\begin{figure}
		\centering
		\includegraphics[width=0.6\textwidth]{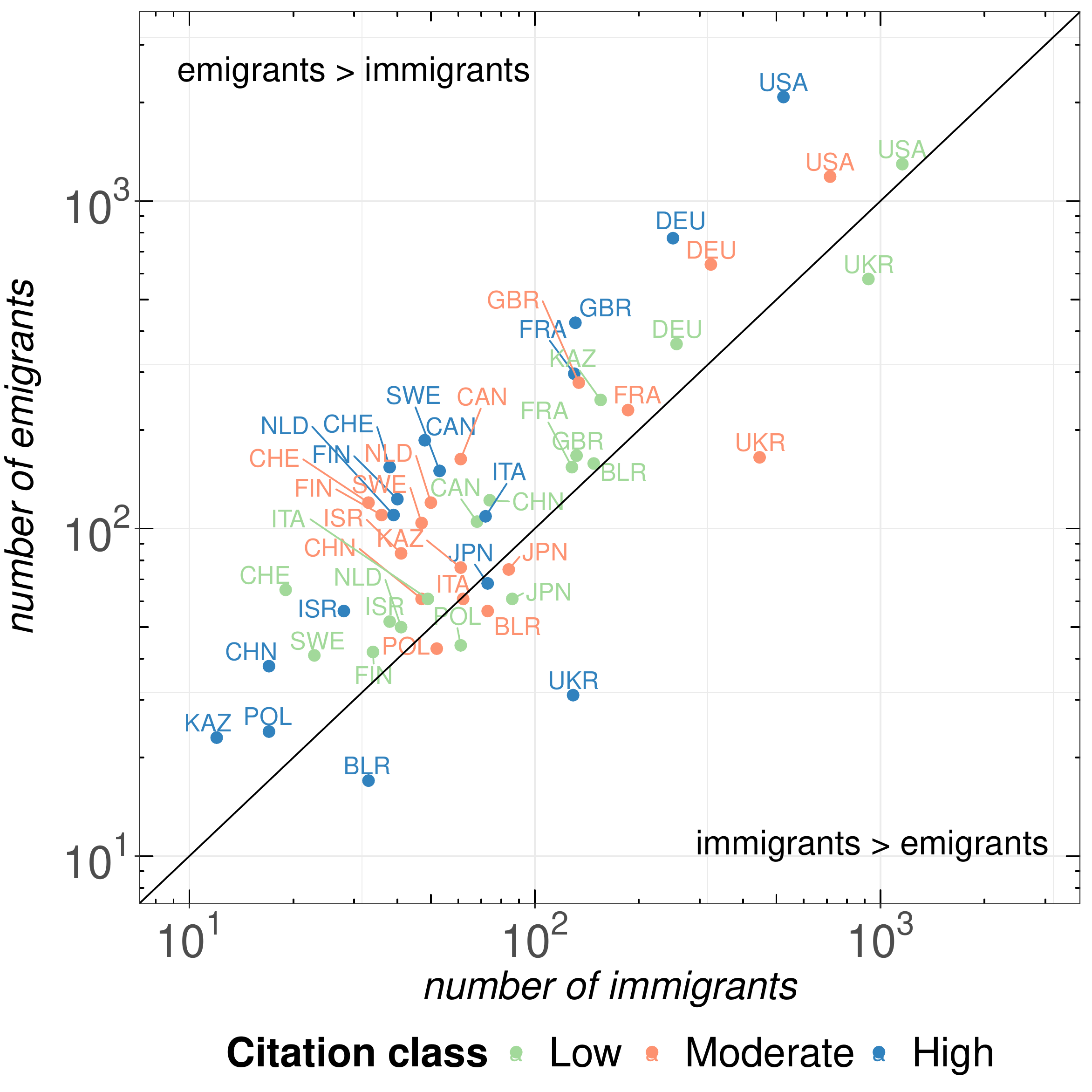}
		\caption{Scatterplot of the number of emigrants by the number of immigrants for each combination of citation class and country. \textit{x} and \textit{y} axes are on a logarithmic scale.} \label{fig:Emi_vs_Imm_cited}
	\end{figure}

	
	To get a better sense of how the number of migrant researchers in each citation class changed based on the direction of the migration flow, we have plotted the numbers of emigrants and immigrants from common countries of origin and destination by citation class in Figure \ref{fig:Emi_vs_Imm_cited}. A general observation that can be made from the log-log scatter plot of Figure \ref{fig:Emi_vs_Imm_cited} is that the higher the citation class is, the greater the disparity is between the number of emigrants and immigrants (green dots are closer to the $45^{\circ}$ line). Except for Ukraine, Japan, Belarus, and Poland, emigrants from Russia to each country outnumbered immigrants in the opposite direction for most citation categories.
	
	\subsection{Net migration rates}
	
	Migration rates are commonly used measures of the difference between movements into and out of a certain area \citep{lieberson_interpretation_1980}. The net migration rate for a given area refers to the difference between in-migration and out-migration rates per 1,000 people. A positive value means that more people are entering than leaving a given area during a certain time period. Using $I_y$ and $E_y$ to represent the number of published researchers who immigrated to Russia and emigrated from Russia, respectively, during year $y$, and $M_y$ to represent the estimated population of scholars in Russia in year $y$, the net migration rate $NMR_y$ can be calculated according to Eq.\ \ref{eq:NMR}. In-migration and out-migration rates can be computed based on $I_y/M_y$ and $E_y/M_y$, respectively, which only concern one direction of the flows. In Eq.\ \ref{eq:NMR}, the denominator, $M_y$, is obtained from all the bibliometric records (of researchers with ties to Russia), which include non-movers as well. It estimates the average number of researchers in Russia in year $y$ based on the mode countries of affiliation associated with publication dates within a two-year vicinity of year $y$. For this estimation, we assume that the researchers with Russian mode countries were in the country two years before and two years after the publication year, unless there is evidence to the contrary (publications showing other mode countries for the researcher). 
	
	\begin{equation}
		\label{eq:NMR}
		NMR_y = {(I_y - E_y)}\times1000/{(M_y)}
	\end{equation}
	\label{ss:rate}
	
\begin{figure}
		\centering
		\subfloat[Net migration rate]{
			\includegraphics[width=0.6\textwidth]{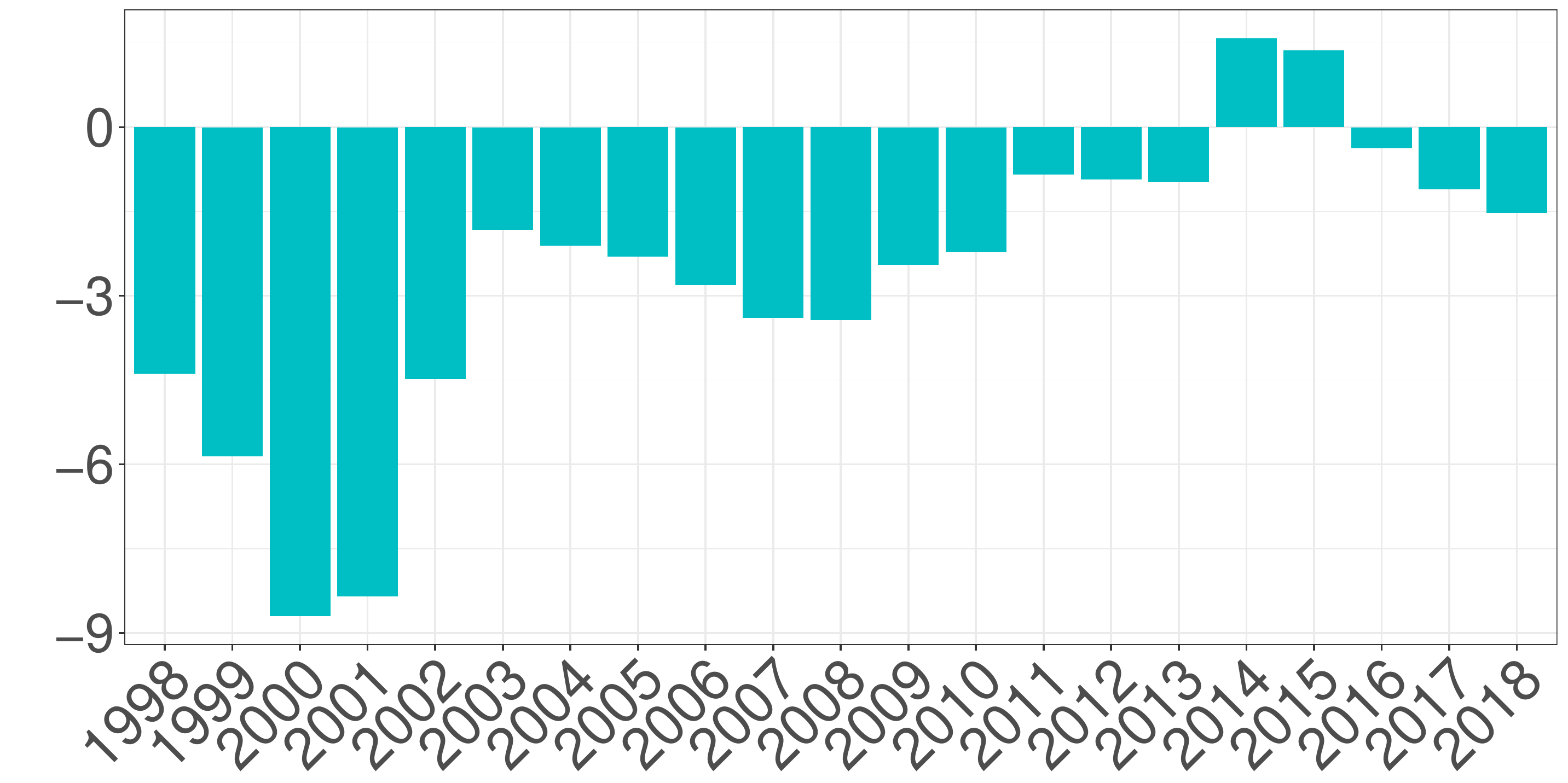}
			\label{subfig:NMR}
		} \hfill
		\subfloat[In-migration rate]{
			\includegraphics[width=0.45\textwidth]{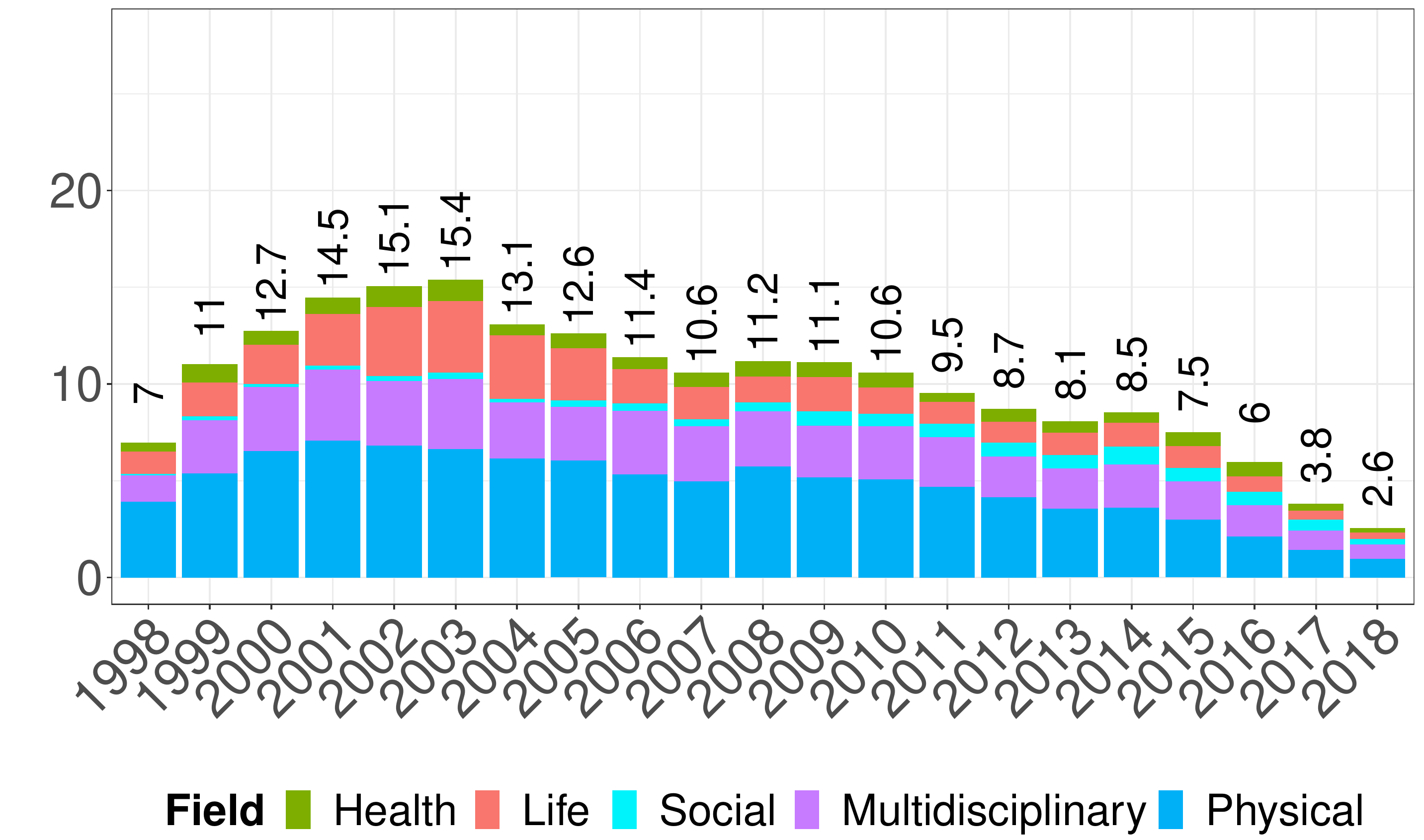}
			\label{subfig:in_migration_rate}
		} \hfill
		\subfloat[Out-migration rate]{
			\includegraphics[width=0.45\textwidth]{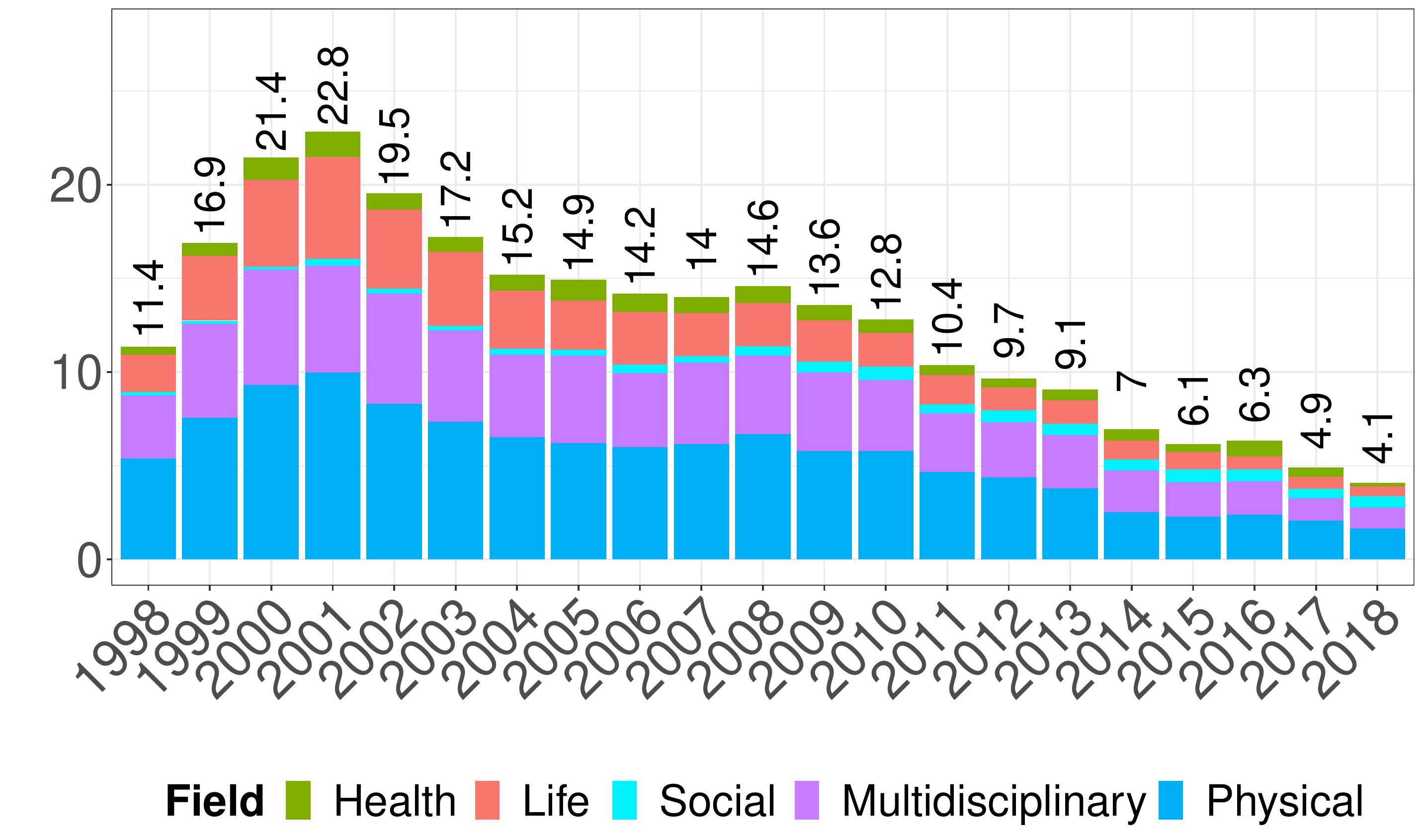}
			\label{subfig:out_migration_rate}
		}
		\caption{Net migration (a), in-migration (b) and out-migration (c) rates per 1,000 researchers in Russia over the 1998-2018 period} \label{fig:countries_of}
	\end{figure}

	Subfigure \ref{subfig:NMR} illustrates the net migration rates in Russia over the 1998-2018 period. The lowest value of the net migration rate is observed for 2000, at $-8.7$ per 1,000 researchers. Looking at the in-migration rate in Subfigure \ref{subfig:in_migration_rate} and the out-migration rate in Subfigure \ref{subfig:out_migration_rate} for 2000, we can see that per $1,000$ published researchers in Russia, $12.7$ published researchers had migrated to Russia, and $21.4$ published researchers had left Russia, resulting in a negative net flow of $8.7$ researchers from Russia to other countries in 2000. Subfigure \ref{subfig:NMR} shows that the net migration rate generally increased over time. The highest rate was in 2014, at $+1.6$. From 2014 onward, we can see a slow downward trend in the net migration rate that ended with the value $-1.5$ in 2018. Subfigures \ref{subfig:in_migration_rate}-\ref{subfig:out_migration_rate} show the shares of each major field of science in the in- and out-migration rates. These findings suggest that the composition of emigrants and immigrants was largely similar, with researchers in the physical sciences being the most common.
	
	\subsection{Overall impact of migration on each discipline}
	\label{ss:subject}
	In this subsection, we analyze the subjects of publications of internationally mobile researchers to evaluate the impact of scholarly migration on different disciplines. To do so, we have developed a measure inspired by the net migration rate for each discipline to quantify the extent to which a given discipline in Russia was affected by the imbalance of incoming and outgoing flows. To operationalize this idea, we start with the concepts of normalized contribution and normalized count formulated in Eqs.\ \eqref{eq:NC} and \eqref{eq:normalizedcount}, which were discussed in Subsection \ref{ss:quantify}.
	
	We evaluate the possible losses in each field by looking at the relative difference between the normalized counts of immigrants, return migrants, emigrants, and transients using a parsimonious measure to quantify \textit{field-based net brain drain} ($FNBD_d$) formulated in Eq. \ref{FNBD}. Emigrants and transients are assumed to increase the net drain of a national science system, and therefore have positive coefficients in Eq.\ \ref{FNBD}. In contrast, we apply negative coefficients to immigrants and return migrants in Eq.\ \ref{FNBD}, because these groups of migrants are assumed to decrease the net drain of a national science system. A larger positive value of $FNBD_d$ means a larger loss due to the imbalance of migration flows in discipline $d$. The largest (smallest) value possible for $FNBD_d$ is $1$ ($-1$), associated with a hypothetical situation in which all migrants in discipline $d$ are emigrants and transients (immigrants and return migrants), the brain drain (brain gain) is therefore at its peak. Each term of Eq.\ \ref{FNBD} represents the impact of the respective group of migrants.
	
	\begin{equation}
		\begin{split}
			\label{FNBD}
			FNBD_d=
			({P^{\text{emi}}_d}/{P_d}) 
			+ ({P^{\text{tra}}_d}/{P_d}) 
			- ({P^{\text{imm}}_d}/{P_d}) 
			- ({P^{\text{ret}}_d}/{P_d})
		\end{split}
	\end{equation}
	
\begin{figure*}
		\centering
		\includegraphics[width=1\textwidth]{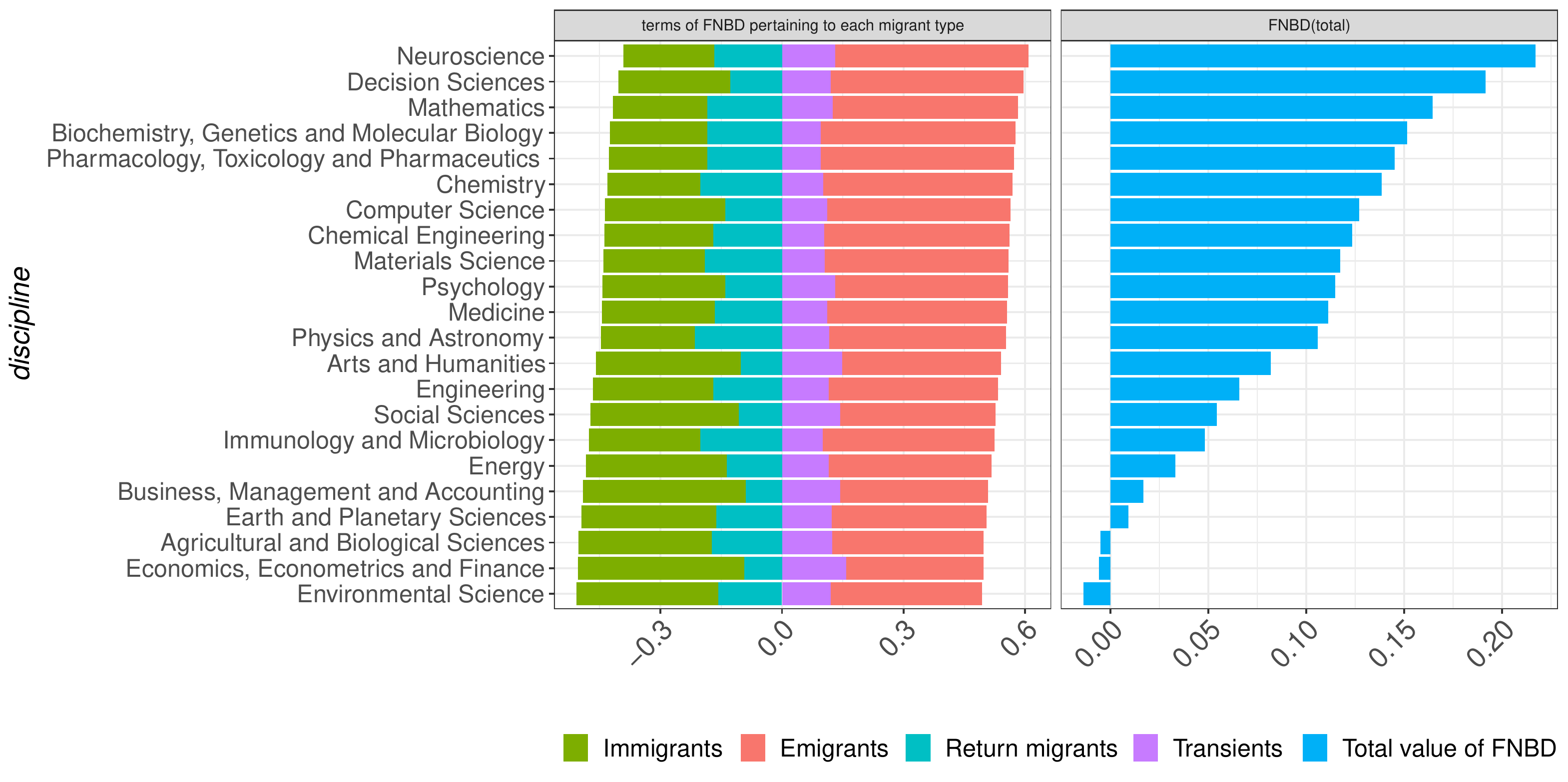}
		\caption{Impact of migration on different fields based on four categories of migrants adding up to a measure of field-based net brain drain}
		\label{ris:fmr_FNBD}
	\end{figure*}

	To illustrate the application of $FNBD_d$, we use the discipline of computer science as an example. We obtain contributions of migrants to different fields using Eq.\ \eqref{eq:NC}, and sum up the normalized contributions in computer science for all four types of migrants (calculating $P^{\text{imm}}_d,P^{\text{emi}}_d,P^{\text{ret}}_d$ and $P^{\text{tra}}_d$) and all migrants together (calculating $P_d$) using Eq.\ \ref{eq:normalizedcount}. Accordingly, the normalized count of mobile researchers in computer science would be $P_d=1511.8$, which includes all four types of migrants. Then, we use the formula in Eq.\ \ref{FNBD} to calculate $FNBD$ for computer science, which is equal to $0.127$. This is interpreted as the overall migration of researchers in Russia over the 1996-2020 period, leading to a $12.7\%$ net drain in the field of computer science. Figure \ref{ris:fmr_FNBD} shows the four terms of $FNBD$ and its total value for 22 disciplines\footnote{Due to the low frequencies of four disciplines in the authorship records of migrants, $FNBD$ values cannot be reliably computed for dentistry, veterinary, nursing, and health professions.}. 
	
	Figure \ref{ris:fmr_FNBD} shows that Russia suffered losses in disciplines such as neuroscience ($21.6\%$), decision sciences ($19.2\%$), mathematics ($16.4\%$), biochemistry ($15.1\%$), pharmacology ($14.8\%$), chemistry ($13.9\%$), computer science ($12.7\%$), chemical engineering ($12.4\%$), materials science ($11.7\%$), psychology ($11.6\%$), medicine ($11.1\%$), and physics ($10.6\%$). For most other disciplines, the $FNBD$ values also show a loss, but to a smaller degree. For some disciplines, the values of $FNBD$ are close to zero, suggesting a relatively balanced circulation of flows when the temporal dimension of the movement is compressed. This is the case for the five disciplines of (1) business, management and accounting ($1.8\%$), (2) earth and planetary sciences ($0.9\%$), (3) agricultural and biological sciences ($-0.5\%$), (4) economics ($-0.7\%$), and (5) environmental science ($-1.4\%$). 
	
	The results indicate that the impacts of the international mobility of researchers on the various fields of science in Russia were heterogeneous. This observation challenges the perspective that a national science system is a single unit with a simple positive/negative response to international mobility. Instead, it appears that the components of such a system could be affected differently by the balance of migration flows, or the lack thereof. Note that if we only consider the terms of $FNBD$ pertaining to immigrants and emigrants, the alternative measurements also show that Russia was suffering a net loss in most disciplines, because emigrants outnumbered immigrants in most disciplines. However, given that large shares academic migrants are return migrants and transients, failing to include them in the analysis could create a distorted picture.
	
	\section{Limitations}
	\label{s:limit}
	A major limitation of this study, as well as a considerable strength, is our use of bibliometric data, and the unique view they provide. It could be argued that some issues associated with bibliometric data are likely to cause not just random noise, but systematic biases. Three examples of potential biases are that (1) women who change their family name could be more likely to be incorrectly issued more than one author ID; (2) authors with a common name could be more likely to suffer from being merged with others; and (3) authors publishing in certain fields could be less likely to have some types of publications in Scopus. We must acknowledge that bibliometric metadata are not produced for use as research data, and like most sources of big data, they are not immune to potential biases or errors. However, within the context of individual-level bibliometric data, Scopus claims to provide greater coverage \citep{scopus_coverage} and higher precision \cite{loktev2020} than its counterparts \citep{falagas2008comparison,mongeon_journal_2016}. With our conservative approach of inferring migration from changes of the mode countries as opposed to all changes of affiliation \citep{robinson-garcia_many_2019}, our results are expected to be less sensitive to issues of precision flaws. In addition, while the precision of Scopus author IDs can be considered high, we used an author disambiguation algorithm to further limit the effects of data quality errors on our numerical results. Future scientometrics studies can advance our understanding of these data quality issues, and propose practical methods for resolving them more effectively.
	
	The time required to conduct and publish research is an important factor \citep{cohen_scholarly_2019} to keep in mind when interpreting the temporal component of the results on mobility patterns observed through bibliometric data. In some cases, it may take years from the initiation to the publication of a research project, and such time lags prevent us from observing the movements of researchers with an ideal level of temporal accuracy. In addition, it should be noted that the one-time usage of an affiliation does not necessarily indicate a direct attachment to the country of affiliation \citep{kosyakov_synchronous_2019}. Our conservative approach tackles this issue by identifying a researcher as an international mover only if the researcher’s mode country of affiliation changed across different years. Moreover, we cannot observe and track migration events that are not represented in publications indexed in Scopus. Bibliometric databases could be biased, and some countries, scientific fields, and languages could be under-represented \citep{falagas2008comparison,mongeon_journal_2016,sugimoto_measuring_2018}. This potential bias exists even though Scopus provides content in 40 languages, and 22\% of the content of Scopus is published in languages other than English \citep{scopus_coverage}. In addition, as we investigated a specific period of time, our data suffers from left-truncation.
	
	Despite these limitations, it is clear that the use of bibliometric data facilitates the study of the migration of researchers across disciplinary boundaries, and allows such analyses to be more current and extensive than would be possible using traditional data sources \citep{alburez2019}. This study makes several contributions to the literature, including methodological contributions through the repurposing of bibliometric data, and substantive contributions to research on scholarly migration in Russia. A missing piece of the puzzle in analyses of academic brain drain has been comprehensive migration statistics, which our study provides for the first time for Russia, a country that is often at the center of brain drain debates, even though quantitative analyses of brain drain in Russia based on large-scale data have seldom been conducted. 
	
	\section{Sensitivity Analysis}
	\label{ss:appendix}
	Some of the challenges of using big data for research are data incompleteness and non-representativeness \citep{salganik_bit_2019}. In particular, bibliometric data have several limitations, which we discussed in Section \ref{s:limit}. Despite these challenges and limitations, large-scale data are sometimes viable alternatives to traditional data, as they allow researchers to improve existing statistics, and to explore issues that remain elusive when using traditional data \citep{weber_digital_2017}. In the context of population studies, migration statistics are usually inconsistent across countries and are often outdated \citep{de_beer_overcoming_2010, zagheni_inferring_2014}, which makes big data on migration particularly useful. In Section \ref{s:results}, we showed the novel insights that large-scale bibliometric records can provide if they are carefully pre-processed and analyzed, while acknowledging and responding to the sources of potential data quality issues that may affect the results.
	
	To evaluate the reliability of our results on migration of scholars, we conduct three sensitivity analyses in this section.
	
	\subsection{Sensitivity analysis on $NMR$}
	
\begin{figure}
		\centering
		\subfloat[Boxplot of absolute variance of NMR indifferent scenarios of randomly excluding data]{
			\includegraphics[width=0.45\textwidth]{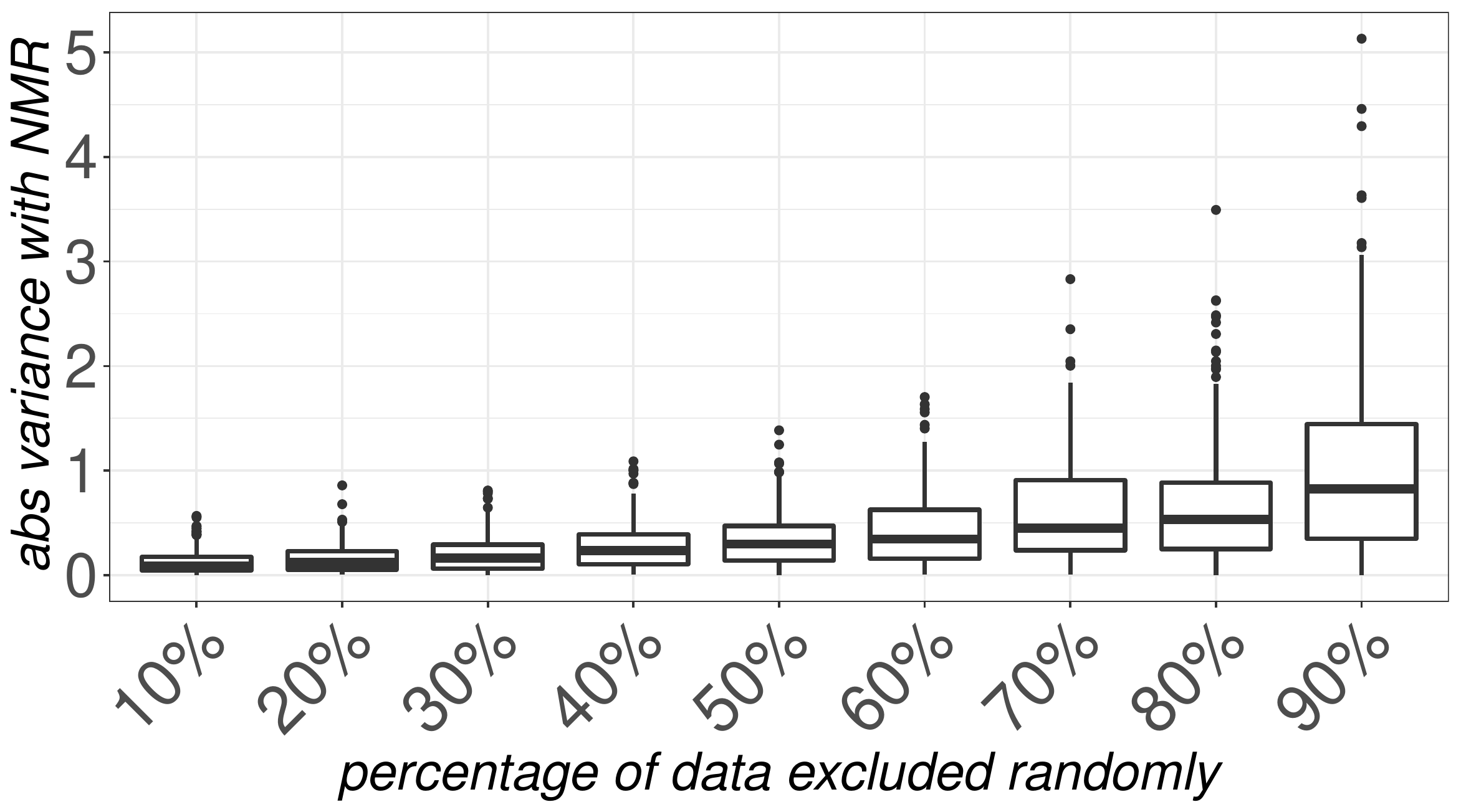}
			\label{subfig:boxplot_abs_var_NMR}
		} 
		\hfill
		\subfloat[Net migration rates for different scenarios of randomly excluding data]{
			\includegraphics[width=0.45\textwidth]{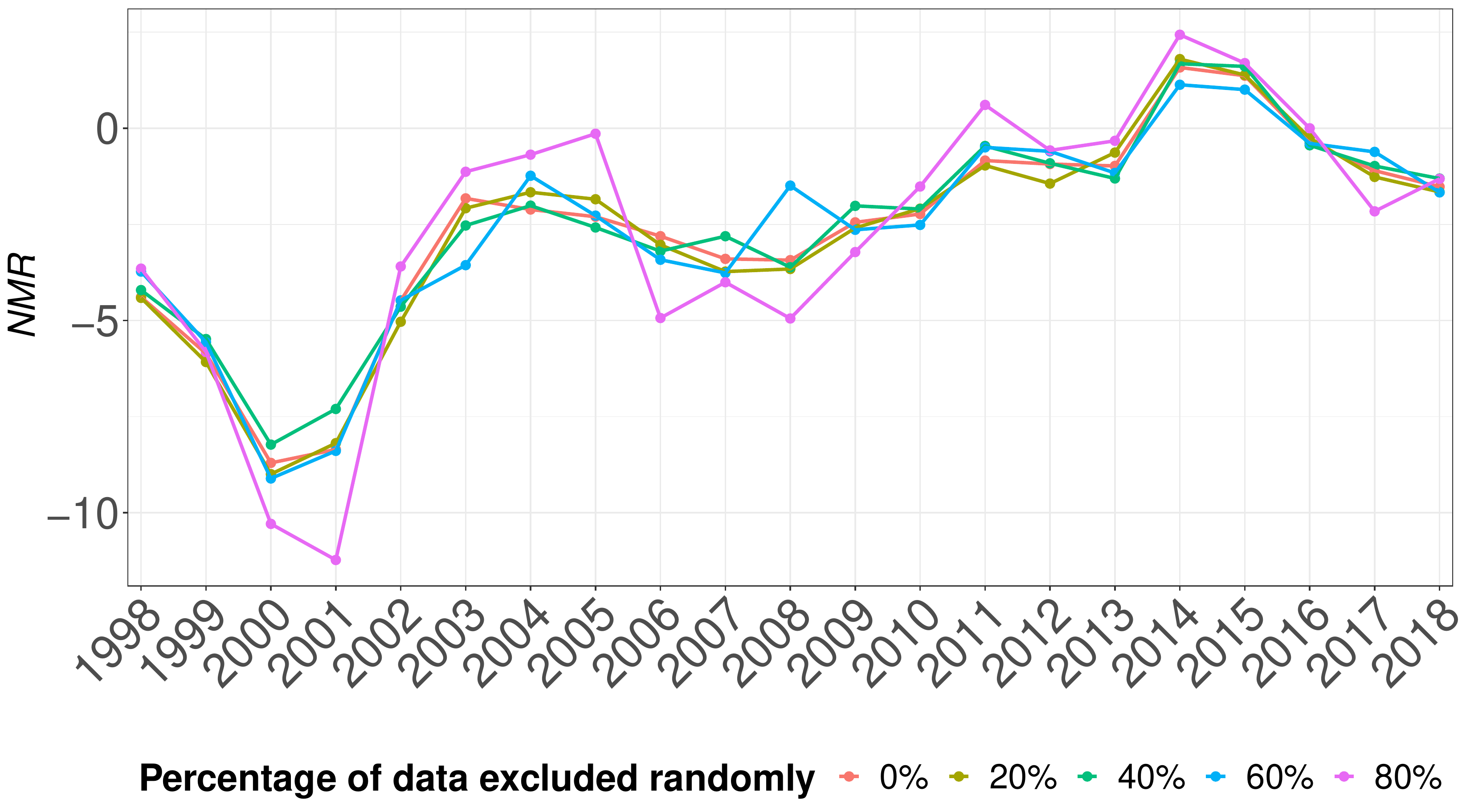}
			\label{subfig:NMR_different}
		} 
		\caption{Sensitivity analysis of net migration rates over the 1998-2018 period showing fairly stable results even when the majority of the data are excluded} \label{fig:sensitive_NMR}
	\end{figure}

	We first conduct a sensitivity analysis in Figure \ref{fig:sensitive_NMR} to check the robustness of our net migration rate estimates. This is done by randomly removing a proportion of the original data from our dataset and investigating the impact of this removal on the numerical results. In Subfigure \ref{subfig:boxplot_abs_var_NMR} we randomly exclude proportions of the data in multiples of $10\%$. We observe the expected result that the less data we have, the more variance there is in the net migration measurements. Subfigure \ref{subfig:NMR_different} shows the net migration rates produced for five scenarios of randomly excluding data in multiples of $20\%$. In Subfigure \ref{subfig:NMR_different}, each line represents a single run. Overall, we can see that the measurements are fairly stable. Even when excluding $80\%$ of the data, our main inference on the temporal trends of the net migration rate (initially declining until 2000, then increasing up to a positive value in 2014, and then declining until 2018) remains robust and stable.
	
	\subsection{Sensitivity analysis on population estimates}
	
	As the denominator for the net migration rate, we used estimates of the population of researchers based on the assumption that researchers published with Russia as the mode country of affiliation were in the country \textit{two} years before and \textit{two} years after the publication year, unless there is evidence to the contrary. As another sensitivity analysis, we change the number of years (the padding parameter) that a researcher is retained in the total count from  \textit{two} to alternative values $\{1,3,4,5\}$.
	
	\begin{figure}
		\centering
		\includegraphics[width=0.48\textwidth]{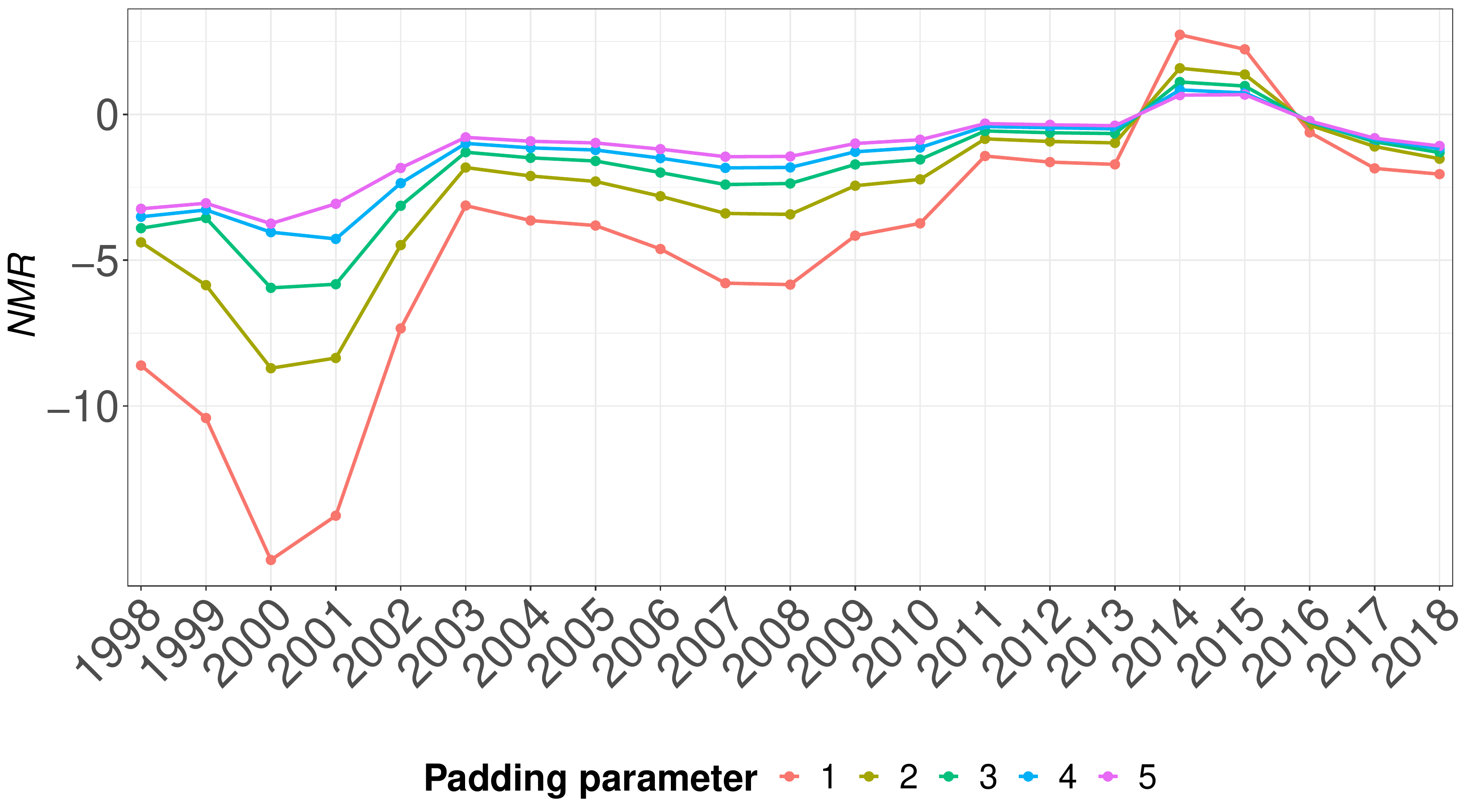}
		\caption{Net migration rates calculated based on different scenarios for the population of scholars in Russia over the 1996-2020 period} \label{fig:NMR_diff_scenarios}
	\end{figure}
	
	Figure \ref{fig:NMR_diff_scenarios} displays different scenarios for the net migration rate of researchers in Russia based on different padding parameter values. The results shown in Figure \ref{fig:NMR_diff_scenarios} indicate that modifying this parameter mainly changes the scale of the net migration rate, while its trend remains unchanged.
	
	\subsection{Sensitivity analysis on $FNBD$}
	
\begin{figure*}
		\centering
		\subfloat[Boxplot of absolute variance of FNBD in different scenarios of randomly excluding proportions of the data]{
			\includegraphics[width=0.98\textwidth]{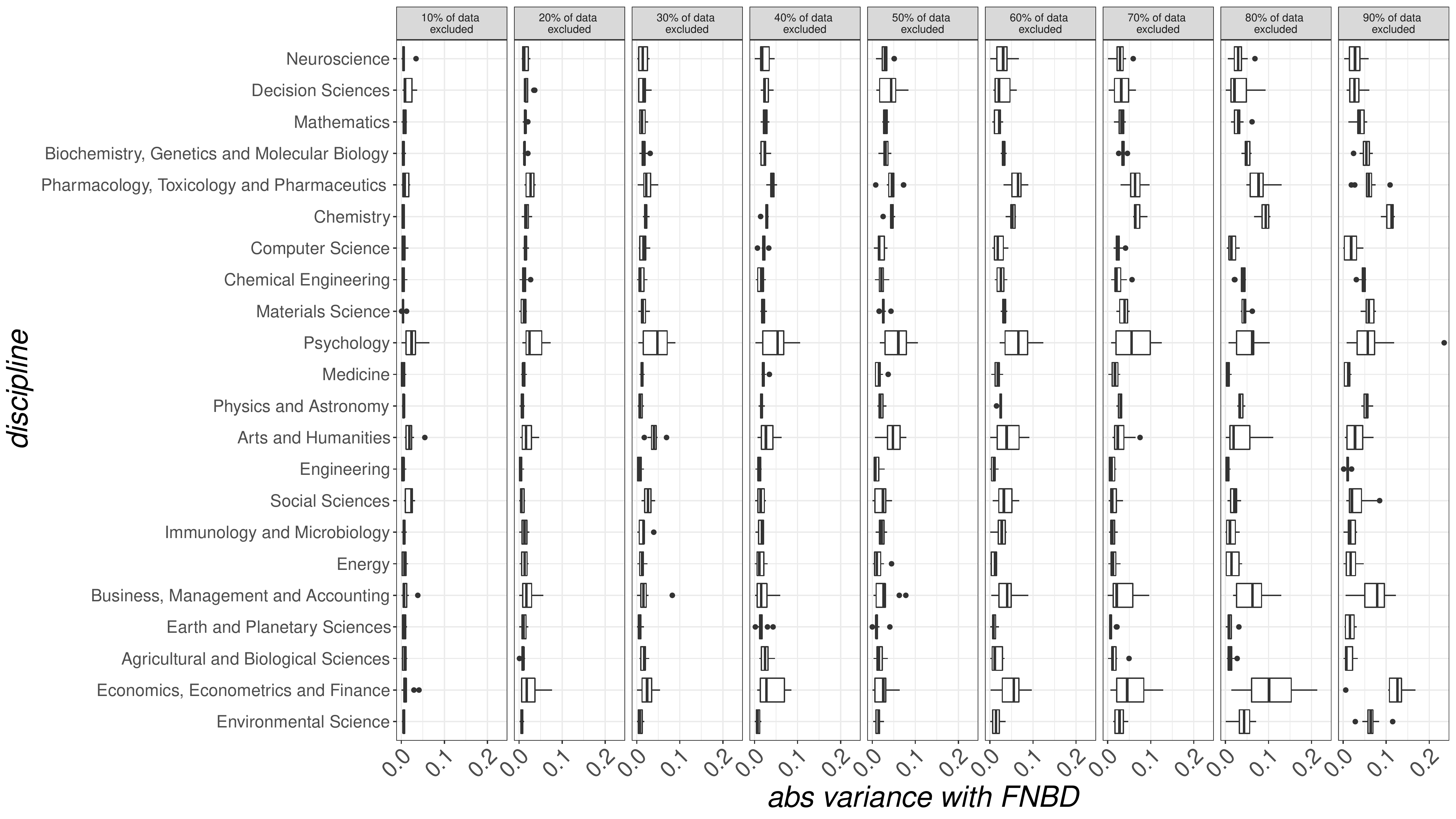}
			\label{subfig:boxplot_abs_var_FNBD}
		} \hfill
		\subfloat[FNBD calculated based on different scenarios of randomly excluding proportions of the data]{
			\includegraphics[width=0.98\textwidth]{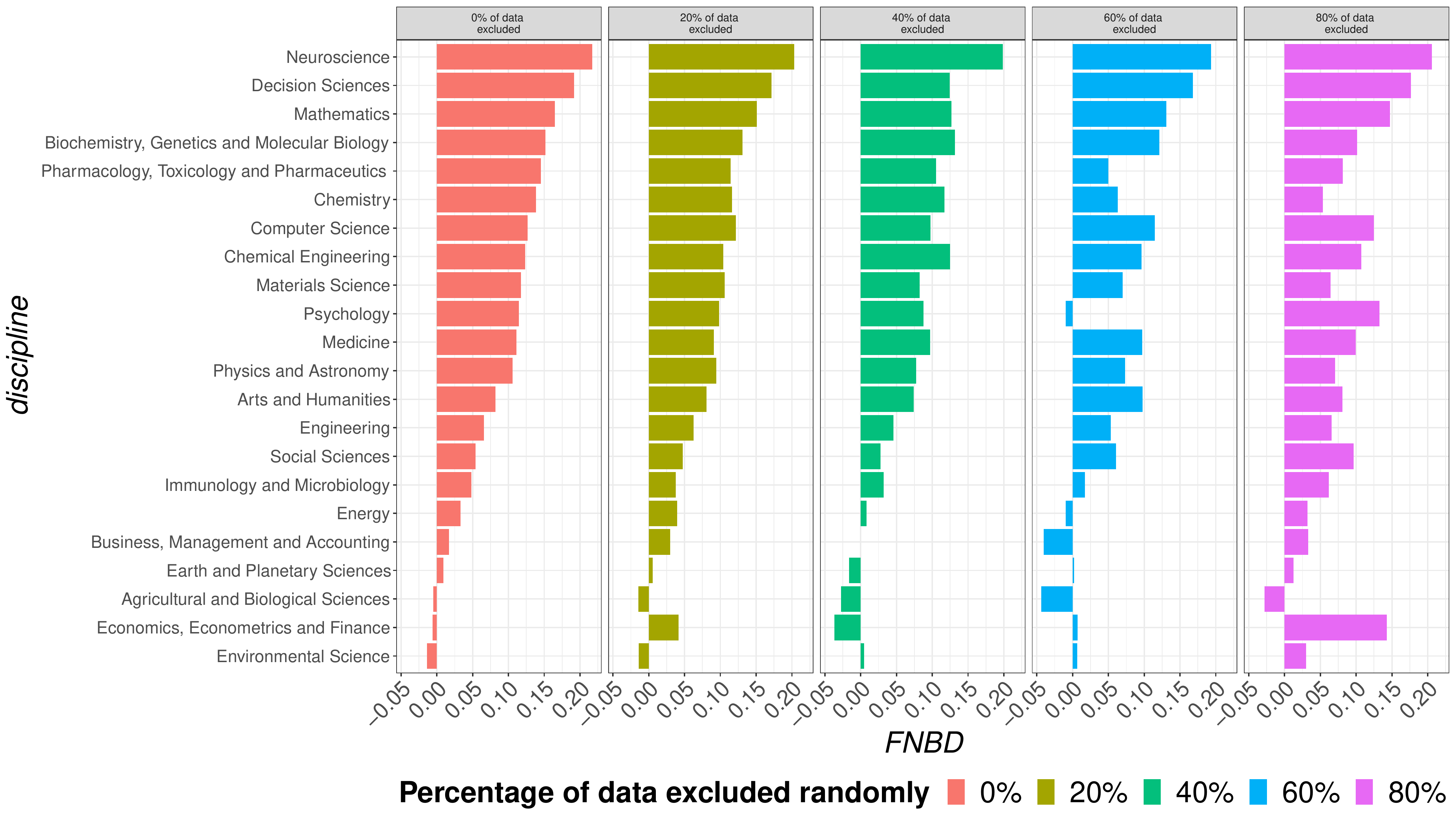}
			\label{subfig:FNBD_different}
		} 
		\caption{Sensitivity analysis of FNBD in Russia over the 1996-2020 period} \label{fig:sensitive_FNBD}
	\end{figure*}

	A similar sensitivity analysis based on data exclusion is performed for our measurements of $FNBD$. In Figure \ref{fig:sensitive_FNBD}, we provide the results of the sensitivity analysis as the boxplot of absolute variance (in Subfigure \ref{subfig:boxplot_abs_var_FNBD}) and alternative values of $FNBD$ for different scenarios of randomly excluding data (in Subfigure \ref{subfig:FNBD_different}). We can see that the measurements for the discipline of psychology are sensitive. A potential explanation for this finding is that this field is particularly uncommon in our dataset.
	
	Although the number of publications in psychology journals by researchers associated with Russia has been growing in recent years, the contributions by Russian scholars still account for less than $1\%$ of the world output in psychology \citep{lovakov_bibliometric_2019}. The sensitivity for the field of economics is also interesting, given that when $20\%$ of the data are excluded, the respective $FNBD$ value becomes positive. This result highlights the sensitivity of the data to a migration measurement that disaggregates the data into small subpopulations. However, the main finding that Russia has been on the losing side in most disciplines does not change even after excluding $80\%$ of the data. When we look at the other disciplines, we can see that while there are some fluctuations, the results are fairly stable, especially when we exclude large proportions of the original data.
	
	
	\section{Discussion and Summary}
	\label{s:discuss}
	In this study, we used affiliation addresses from 2.4 million Scopus publications over the 1996-2020 period in order to present a comprehensive and detailed picture of academic migration in Russia. Our goal was to understand the patterns of scholarly migration by tracking the international movements of researchers, and to determine the impact of such movements on the Russian science system, both overall and in each field of science. The use of large-scale bibliometric data from Scopus allowed us to achieve this goal in a cross-disciplinary study of demography and scientometrics. We documented that while migrant researchers make up only a small minority of all scholars, they play an essential role in the Russian science system, as the substantial differences in the citation-based performance of migrant and non-migrant researchers indicate. We focused on several aspects of this small but highly influential subpopulation of researchers, including their countries of origin and destination, and their disciplines. 
	
	Our analysis of four categories of academic migrants revealed the similarities in their common countries of academic origin and destination, while highlighting the differences in their migration patterns and in their the impact on the Russian science system. The US and Germany were shown to be the largest scientific hubs linked to Russia. Ukraine turned out to be one of the main origins of published researchers who migrated to Russia, which may be explained in part by the patterns of Russo-Ukrainian migration in the general population. Using data on major fields of science, we compared the international flows in major fields of scholarship. The field of physical sciences was found to have the largest flows, followed by the multidisciplinary field. The life sciences field had the third-largest flows, most likely because the technical knowledge and skills associated with this field tend to be internationally transferable. The health sciences field had smaller flows, which may because bodies of knowledge on health tend to vary across countries (e.g., national medical protocols). Finally, the flows were smallest for the social sciences, as research in this field likely depends to a greater extent than in other scientific fields on the language, culture, and context of each society.
	
	We linked the authorship records with Scopus citation data, and observed large disparities in citation-based performance between migrants by their mobility types and across different fields. Consistent with the generally observed pattern in scientometrics \citep{Leydesdorff_citation, marmolejo2015mobility,bedenlier2017internationalization,horta2019mobility}, we found that among migrant researchers with ties to Russia, the physical sciences had the most citations, followed by the life sciences and the health sciences (which are somewhat comparable); whereas the social sciences had the fewest citations. From the perspective of mobility types, emigrants from Russia received far more citations than immigrants to Russia in all fields except for the social sciences, for which immigrants and emigrants received almost equal numbers of citations, on average. The disparities observed for the physical sciences, the life sciences, and the health sciences may be attributed to a multitude of factors, including the research performance of immigrants and emigrants and the differences in the research opportunities in their respective destination countries. For instance, access to laboratories and other equipment and infrastructure in the science, technology, engineering, and mathematics (STEM) fields varies considerably across countries \citep{horta2019mobility,yonezawa_mobility_2016}. The balance of the citation-based performance levels of emigrant and immigrant social scientists may indicate that the consequences of international moves are different in the social sciences than they are in other fields. In the physical sciences, the life sciences, and the health sciences, migrants who returned to their origin countries tended to have higher citation rates, which may indicate that they had more publications, and that their research had received more scientific recognition and international exposure. In contrast, returning social science researchers had a citation rate similar to that of stayers, which may suggest that moving internationally had less pronounced effects on their citation-based performance.
	
	Further grouping the migrant researchers into three categories based on their annual citation rates (normalized by age and field), we compared the numbers of emigrants and immigrants by country, and looked at their citation-based performance. Lowly cited researchers were found to be the dominant category for both immigrants and emigrants in Kazakhstan, Uzbekistan, and Ukraine. By contrast, the US, Germany, the UK, Sweden, France, Italy, and Spain had large shares of moderately cited and highly cited researchers, especially among emigrants. Our analysis also demonstrated that in the higher citation classes, the emigrants from Russia to each country outnumbered the immigrants in the opposite direction by larger margins.
	
	Using the net migration rate, we found that while Russia had, overall, been on the losing side of a brain circulation system in the late 1990s and early 2000s, this trend has shown signs of reversing in recent years. We found that over the past several years, the net migration rate had become positive or was closer to zero, indicating a relative balance between the annual flows of incoming and outgoing researchers. However, while our results appear to show that the incoming and outgoing flows of researchers have been more balanced in Russia in recent years, it is still too early to call Russia a country of attraction for researchers.
	
	When analyzing the disciplines of internationally mobile researchers, we normalized their contributions in different fields, and introduced a measure of net brain drain to quantify the impact of migration on each specific subfield of scholarship. The results of the analysis showed that over the 1996-2020 time period, there was a relatively large outflow of specialists in most fields of science in Russia. Specifically, our findings indicate that there was a net loss of published researchers for Russia in the fields of neuroscience, decision sciences, mathematics, biochemistry, pharmacology, chemistry, computer science, chemical engineering, materials science, psychology, medicine, and physics. Thus, our results are in line with previous findings about the migration of specialists in these fields \citep{volz_utechka_2002,ball_russian_2005,rybakovsky_international_2005,ryazantsev_russia_2013,antoshchuk_female_2018}, and go a few steps further by providing a comprehensive picture of all mobility types and 22 subfields of science. It is important to point out that these losses are partly attributable to the compression of the temporal dimension of the data, and to the relatively large numbers of specialists who left Russia in the late 1990s and early 2000s (see negative net migration rates for the years 1998-2009 in Subfigure \ref{subfig:NMR}).
	
	The results that this research provided for the first time are generalizable within the limitations of bibliometric data, which were discussed in Section \ref{s:limit}. Keeping some caveats in mind, our substantive and methodological contributions can be used to further our understanding of international migration in academia. For the specific topic of this study, timely and detailed statistics have been difficult to obtain through traditional data sources. Our findings provide new insights for Russia that may be used in the development of high-skilled migration policy. The methodological contributions of this study can be applied to other countries as a framework of analysis for examining the effects of scholarly migration on other national science systems.
	
	We aim to continue this line of research by investigating additional dimensions of analysis, such as gender, and using dynamic measurements of individuals’ characteristics. These approaches will allow for a deeper analysis across disciplines using several indicators of research performance. This study has only scratched the surface of the topic of scholarly migration, and of new applications of bibliometric data. Thus, while many questions remain unresolved, we hope to have paved the way to addressing them.
	
	\section*{Declaration}
	The authors declare that there are no conflicts of interest.
	
	\section*{Acknowledgments}
	The authors highly appreciate the technical support from Tom Theile, and the discussions with Daniela Perrotta, Xinyi Zhao, Andr\'{e} Grow, Ole Hexel, Ebru Sanliturk, Ugo Basellini, Carolina Coimbra, Emanuele Del Fava, Lucia Chen, and Emilio Zagheni that helped in improving the article. The authors are also grateful for the comments from nine anonymous reviewers on earlier versions of this article, which further strengthened the current article. This study has received access to the bibliometric data through the project ``Kompetenzzentrum Bibliometrie", and the authors acknowledge their funder Bundesministerium für Bildung und Forschung (funding identification number 01PQ17001).

\end{document}